\documentclass{aa}  
\usepackage{graphicx}
\usepackage{txfonts}
\begin{document} 

   \title{The multi-wavelength phase curves of small bodies}

   \subtitle{Phase coloring}

   \author{Alvaro Alvarez-Candal
          \inst{1,2}
          }

   \institute{Instituto de Astrof\'isica de Andaluc\'ia, CSIC, Apt 3004, E18080 Granada, Spain\\
              \email{varobes@gmail.com}
        \and
        Instituto de F\'isica Aplicada a las Ciencias y las Tecnolog\'ias, Universidad de Alicante, San Vicent del Raspeig, E03080, Alicante, Spain
             }

   \date{Received XX; accepted XX}

 
  \abstract
  {{Small bodies change their brightness due to different motives: Rotation along their axis or axes, combined with irregular shapes and/or changing surface properties,}
  or changes in the geometry of observations. In this work, we tackle the problem of {\it Phase curves}, which show the change in brightness due to changes in the fraction of illuminated surface as seen by the observer.}
  {We aim to study the effect of the phase curves in the five wavelengths of the Sloan Digital Sky Survey in scores of objects (several tens of thousands), focusing particularly on the spectral slopes and the colors and their changes with phase angle.}
  {We used a Bayesian inference method and Monte Carlo techniques to retrieve the absolute magnitudes in five wavelengths, using the results to study the phase coloring effect in different bins of the semi-major axis.}
  {We obtained absolute magnitudes in the five filters for over 40\,000 objects. Although some outliers are identified, most of the usual color-color space is recovered by the data presented. We also detect a {dual} behavior in the spectral slopes, with a change at $\alpha\sim5$ deg.}
  {}

   \keywords{Methods: data analysis -- Catalogs -- Minor planets, asteroids: general
}
   \titlerunning{Multi-wavelength phase curves of asteroids}
   \authorrunning{Alvarez-Candal}

   \maketitle
%

\section{Introduction}\label{sec:intro}

Large photometric surveys produce (and will produce) massive amounts of data, challenging us to develop new and better methods to make the most of their data products. Among the many surveys ongoing and to come, we would like to call attention to two: The Sloan Digital Sky Survey \citep[SDSS,][]{york2000AJ} and the Legacy Survey of Space and Time \citep[LSST,][]{ivezic2019ApJ} to be carried out by the Vera C. Rubin Observatory. The first one has been very productive for the small bodies community, while the second one will probably challenge all we know about small bodies due to the large number of objects to be observed and the depth of the survey {\citep[$r'=24.7$ for a single visit, see][]{jones2009lsst,ivezic2019ApJ}}. Both surveys use similar photometric systems, the {\it ugriz} system \citep{fukugita1996AJ} increasing the synergy between both datasets and with other surveys using similar photometric systems, for instance SkyMapper \citep{wolf2018SM}, or different wavelength coverage, like the near-infrared \citep{popescu2016movis,carry2018Aeuclid}.

Among the many science cases explored with the SDSS data, we will focus on small bodies' phase curves (PC) to study the phase coloring effect. {A PC relates the change of apparent brightness, normalized to unit distances from the Sun and the observer, of a small body with the phase angle, $\alpha$,} and provides information regarding the micro and macro properties of the surfaces, as well as its absolute magnitude ($H$) {\citep[e.g.,][]{Nelson1998Icar,shkuratov2002Icar,Grynko2008}}. The absolute magnitude in the V filter of a small body is defined as the object's apparent magnitude if illuminated by the Sun at 1 AU while being observed from a distance of 1 AU at opposition (i.e., $\alpha=0$ deg). $H_V$ is related to the size ($D$) and albedo ($p_V$) of the object through the well-known relation \citep{bowelllumme1979}:
\begin{equation}\label{eq1}
    D~[km]=1.329\frac{10^{(3-H_V/5)}}{p_V^{0.5}}.
\end{equation}

Taking advantage of the serendipitous observations of small bodies in the SDSS data at different epochs and orbital configurations, \cite{alcan2022} developed a method to analyze the apparent magnitudes, together with the ephemeris at the time of observations, in particular topo- and heliocentric distances, $\Delta$ and $R$, both in astronomical units (AU) and phase angle ($\alpha$). The method produces multi-wavelength PCs and absolute magnitudes in all the filter system $(H_{\lambda})$. In this work, we further improve the algorithm and apply it {to a different set of measurements derived from the SDSS survey}, as described below.

The work we present here goes in line with a series of works analyzing massive databases, using different methods: \cite{oszki2011pcs} made a comprehensive study of the Minor Planets Center database, \cite{veres2015} analyzed the Pan-STARRS dataset producing about a quarter of a million absolute magnitudes. More recently, \cite{mahlke2021} used ATLAS data to produce two-wavelength PCs, while \cite{colazo2021} used GAIA (DR2) data to obtain $H_G$ for over 10\,000 objects. They all intend to work with large databases, but they have yet to attempt a consistent analysis of the phase coloring along the Solar System.

Phase reddening, which we have been calling phase coloring until now, has been recognized as an important effect that may change the colors or the spectral slope of the objects \citep[for example][]{Taylor1971AJ,Millis1976,luujewitt1990AJ}. It has also been shown that it may impact the taxonomic classification of objects in the border between taxa \citep{sanchez2012}. A clear understanding of the phase reddening effect is crucial for studying the space weathering acting on the small bodies' surfaces because it has a similar effect, and disentangling both is not straightforward \citep{sanchez2012}. Finally, in this paper, we argue that the term phase {\it reddening} should be changed to the more descriptive phase {\it coloring}, terminology that will be used in the remaining text.

This paper is structured as follows:
Description of the data in Sect. \ref{sect:data}, while the improvements made to the algorithm are described in Sect. \ref{sec:analysis}. The results are presented and discussed in Sect. \ref{sec:discussion} to draw the Conclusions in the last section.

\section{Dataset}\label{sect:data}

We use the data of small bodies extracted from the SDSS, but not the traditional and well-known Moving Objects Catalog \citep{ivezic2001AJ,juric2002AJ} frozen in its fourth release, with data obtained until March 2007. Instead, we use the re-analysis of the SDSS performed by \cite{sergeyev2021AyA}, hereafter S21, who released more than one million observations of almost 380\,000 small bodies, multiplying by three the number of objects with respect to the last release of the Moving Objects Catalog.

\begin{figure}
\centering
\includegraphics[width=\hsize]{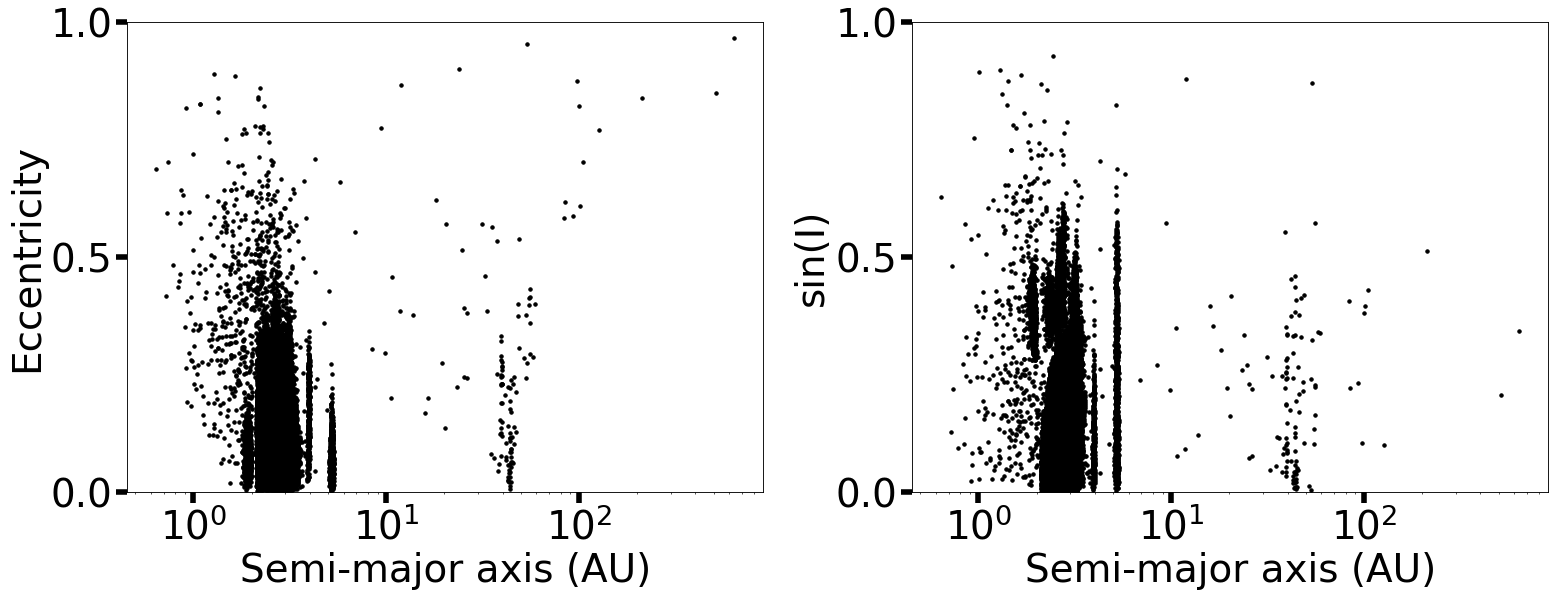}
\caption{Osculating orbital elements of all objects analyzed in this work with at least one $H$ computed in any filter.}\label{fig:orbelem}%
\end{figure}

{The S21 catalog includes the point spread function magnitudes in the ugriz filter system and their respective uncertainties, $\sigma_m$. We will use {\it italics} when discussing magnitudes in the text. S21 also includes the heliocentric and topocentric distances of the objects and the phase angle (along with other helpful information).} As in \cite{alcan2022}, AC22 hereafter, we do not make any cut in the database other than removing {any $m$ with an associated error of $\sigma_{m}\geq1$}. We use data of objects with at least three different $\alpha$ and with a minimum span of $\Delta\alpha\geq5.0$ degrees (i.e., the distance between the maximum and the minimum $\alpha$). Figure \ref{fig:orbelem} shows the distribution in the orbital elements space of all objects with at least one $H$ measured in any filter (see below).

To analyze the data, we built upon the algorithm developed in AC22, with a few modifications intended to improve accuracy, as discussed in the next section.

\section{Analysis}\label{sec:analysis}

We use the method described in AC22, updated as described below. The HG$_{12}^*$ model \citep{penti2016HG} is used as the photometric model because it proved to be better suited to fit the data {and because alternative models presented different problems when applied to the S21 dataset}. The HG$_1$G$_2$ model \citep{muinonen2010HG1G2} needs well-sampled phase curve, {which is usually not the case of the data analyzed in this work}, the Shevchenko model \citep{belskayashev2000} needs a good coverage of the phase curve in small phase angles; and, the HG model \citep{bowell1989HG} fails to represent phase curves of high or low-albedo asteroids. {In this case, we did not use the Python adaptation of the HG$_{12}^*$ model provided by the authors} but a code based on the equations presented in \cite{penti2016HG}.

The adopted photometric model describes the phase curves as follows: 
\begin{equation}\label{eq:2}
M(\alpha)=H-2.5\log{[G_1\Phi_1(\alpha)+G_2\Phi_2(\alpha)+(1-G_1-G_2)\Phi_3(\alpha)]},
\end{equation} 
where $M(\alpha)=m-5\log{(R\Delta)}$ is the reduced magnitude, $H$ is the absolute magnitude, $\Phi_i$ are known functions of $\alpha$ (see \citealt{penti2016HG} for the tables), and $G_i$ are the phase coefficients that provide the shape of the curve. In the HG$_{12}^*$ model, $G_1$ and $G_2$ are parameterized by $G_{12}^*$ according to $G_1=0.84293649G_{12}^*$ and $G_2=0.53513350(1-G_{12}^*)$. From now on, $G_{12}^*$ is $G$ for simplicity unless explicitly stated otherwise, and we will not use sub-indexes to differentiate among filters unless necessary.

\begin{figure}
\centering
\includegraphics[width=4.4cm]{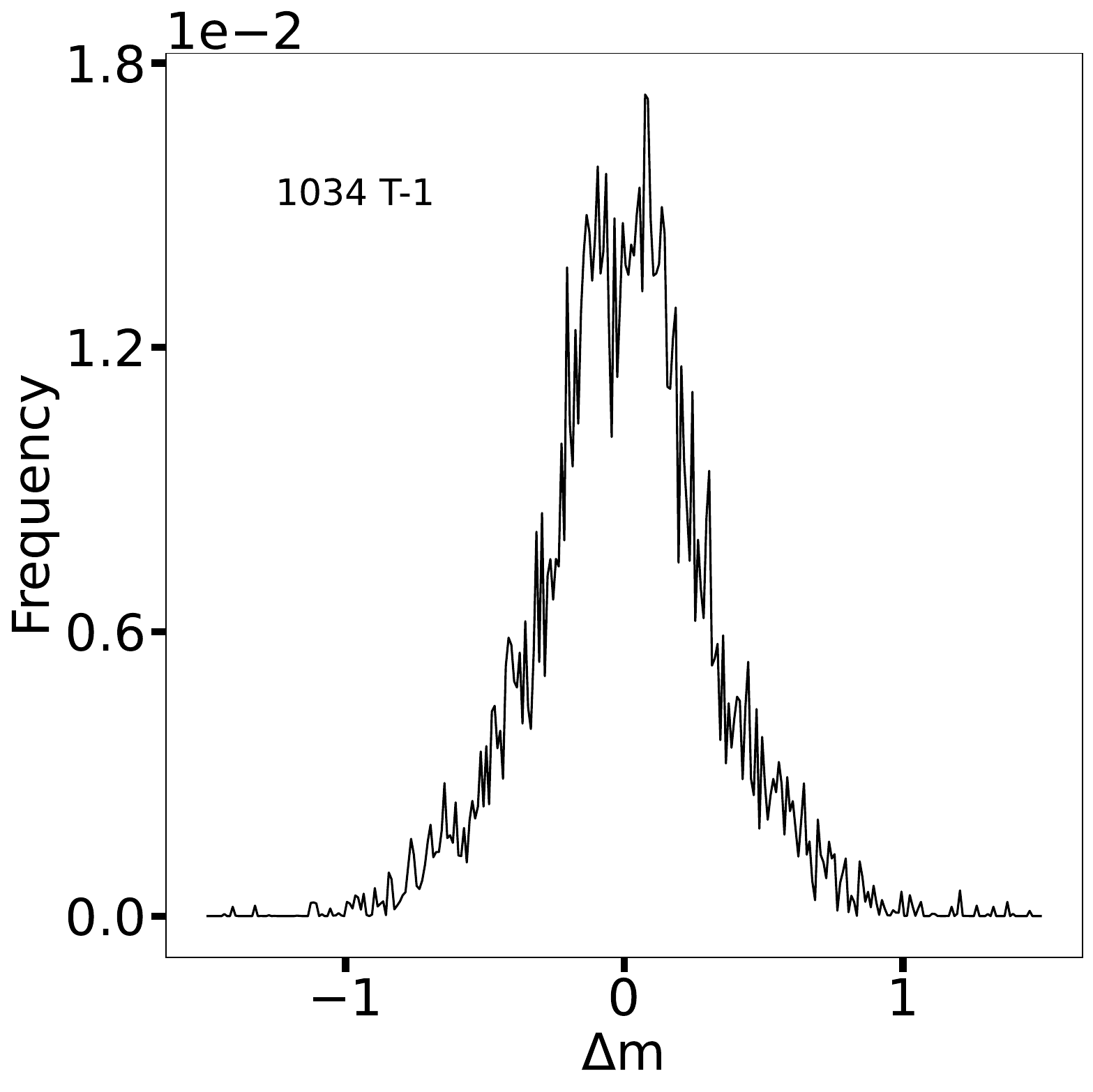}
\includegraphics[width=4.4cm]{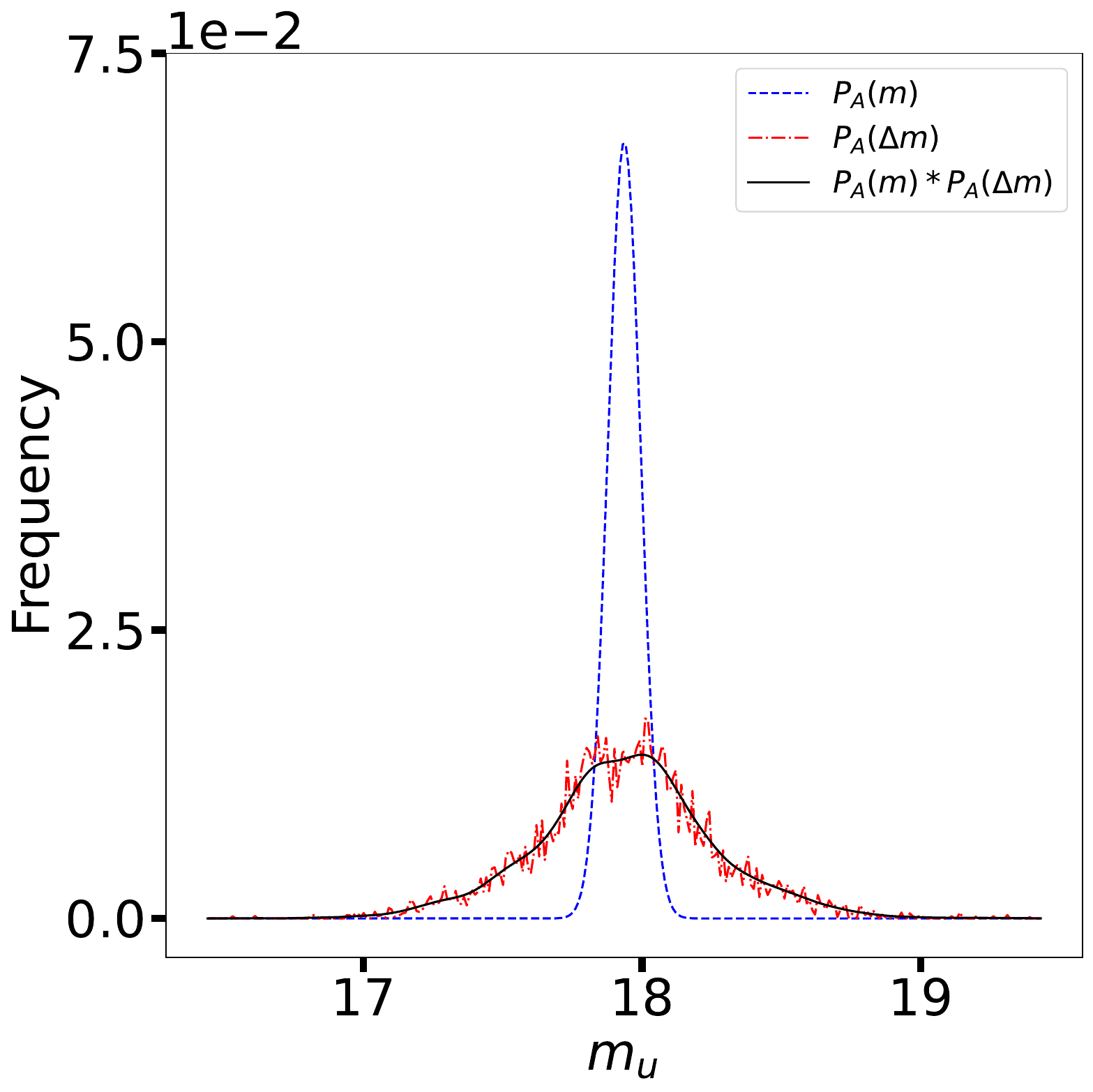}
\caption{Probability distributions. Left: Probability distribution  $P_{29030}(\Delta m)$. Right: Example of the final probability distribution. The examples are for  asteroid 1034 T-1 or 29030 (MPC designation) with $m_v=17.93\pm0.06$.}\label{fig:pdfdm}%
\end{figure}

\paragraph{Probability distribution function, PDF, of $\Delta m$}
We followed a similar approach as in AC22, with one modification. \cite{alcan2022} did not include any effect due to the possible rotational phase at which the objects could have been observed; this produced an artifact, a dip, in $P_A(\Delta m)$ as seen in Fig. 2 of AC22. The dip is alleviated when convolving with the errors in the apparent magnitudes, but it still assigned less probability than real to the region around $\Delta m = 0$. We changed the code to account for this by multiplying a random $\sin(\phi)$, where $\phi\in[0,2\pi)$, simulating an observation at an unknown rotational phase. Equation 2 in AC22 is then re-written as
\begin{equation}\label{eq:3}
P_A^i(\Delta m)=
        \frac{\sum_j{P^i(\Delta m|P_j(H_{V}))\sin(\phi_j) P_j(H_{V})}}
             {\sum_i{\sum_j{P^i(\Delta m|P_j(H_{V}))\sin(\phi_j) P_j(H_{V})}}}
             ,
\end{equation}
where $P_A(\Delta m)=\{P_A^i(\Delta m)\}$ is the PDF of $\Delta m$ for object $A$ in 150 bins of width 0.02 mag, meant to cover the whole range of reported $\Delta m\in[0.0,3.0]$ \citep{wagner2019}. The denominator is a normalization factor. With this procedure, we created a PDF of possible values of $\Delta m$ that the object could have (see an example in Fig. \ref{fig:pdfdm} for asteroid 29030\footnote{The asteroid 29030, or 1034 T-1, is used as an example because it was already used in AC22 and \cite{alvcan2022}.}).

The rest of the process is the same as in AC22. Quoting: 
``To create the final PDF used to estimate the final range of possible solutions of $H$ in all filters, we must also include the uncertainty in the measurement of $m$. In this case, we construct for each object and measure the distribution, $P_A(m),$ as a normal distribution with $\sigma=\sigma_m$ centered at $m$. The final PDF is the convolution of $P_A(m)$ and Eq. \ref{eq:3}:
\begin{equation}\label{eq:4}
    P_A(m,\Delta m)=\big(P_A(m)*P_A(\Delta m)\big).{\rm ''}
\end{equation}
One vantage of including a factor $\sin{(\phi)}$ in Eq. \ref{eq:3} is that it is not necessary to mirror and re-scale $P_A(m,\Delta m)$ as AC22 needed because now the distribution naturally covers the whole space of possible $\Delta m$.

\paragraph{Computation of the phase curve}
As in AC22, we compute $H$ and $G$ for each object and filter using Eq. \ref{eq:2} generating different solutions of the PCs extracting 10\,000 values randomly from $P_A(m,\Delta m)$, the method is explained in full in AC22.

The outcome of the processing is the PDF of $H_{\lambda}$ and their corresponding $G$s if there are at least 100 valid, {that is, physically possible} solutions. All PDFs are available for the readers upon request or in the Open Science Framework (OSF) project folder {\tt https://osf.io/43eny/}. Table \ref{table:1} shows a sample of the nominal values of the quantities and their uncertainties. The former is the median of the PDFs, while the uncertainty range indicates the 16th and 84th percentiles. The complete table is available online, but without format, in the OSF folder and will be available and formatted at the CDS upon acceptance.
\begin{table*}
\caption{Sample of the catalog.}
\label{table:1}
\centering
\begin{tabular}{c| c c c c |c c c | c c }
\hline\hline
ID-MPC designation & $H_u$ & $H_u^-$ & $H_u^+$ & N & $G_u$ & $G_u^-$ & $G_u^+$ & $\alpha_{min}$ (deg) & $\Delta\alpha$ (deg) \\
\hline
1013 T-2 - 48357&19.271 &  19.502 &  19.029 & 6 &   0.913 &   0.349 &  1.336 &   7.42 & 16.19\\
1027 T-2 - C9386&17.719 &  17.916 &  17.502 & 4 &   0.839 &   0.286 &  1.297 &   2.22 & 12.65\\
1033 T-2 - B7991&-8.0   &  -8.0   &  -8.0   & 5 &   -8.0  &   -8.0  &  -8.0  &  12.70 &   4.8\\
1034 T-1 - 29030&17.491 &  17.587 &  17.400 & 5 &   0.445 &   0.135 &  0.934 &  16.30 & 22.44\\ 
1037 T-3 - B8086&18.523 &  18.790 &  18.310 &17 &   0.778 &   0.267 &  1.259 &   9.67 & 18.12\\
\hline
\end{tabular}
\tablefoot{The first column shows the ID of the object and its Minor Planets Center ASCII designation. $H_u$ is the median of the probability distribution, while $^-$ and $^+$ designate the absolute magnitude at the 16th and 84th percentiles. The number of observations considered is shown in the fifth column. $G_u$, $G_u^-$, and $G_u^+$  from column six to eight. The last two columns show the minimum $\alpha$ and its total span. Flag -8 indicates objects with less than three observations in a given filter or $\Delta\alpha<5$ deg, while flag -9 indicates that not enough valid solutions were obtained {(that is, less than 100)}. The complete catalog is available at the CDS (https://cdsarc.u-strasbg.fr/) or upon request.}
\end{table*}

\section{Results and discussion}\label{sec:discussion}

Following the above procedure, we computed $H$ in all filters available in the S21 data (see numbers in Table \ref{table:2}) and also in the $H_V$ and $H_R$ using $g$, $r$, and $i$ for each $\alpha$ when the magnitudes were available. The data was processed the same way as all the rest using $V$ and $R$ and their corresponding $\alpha$. 
\begin{table}
\caption{Number of absolute magnitudes obtained}
\label{table:2}
\centering
\begin{tabular}{c c | c c}
\hline\hline
Magnitude  & N         & {Criteria} & N \\
\hline
$H_u$   & 61\,711   & at least one valid $H_{ugriz}$               & 83\,451 \\
$H_g$   & 61\,753   & valid $H_u$, $H_g$, $H_r$, $H_i$, and $H_z$  & 42\,168 \\
$H_r$   & 80\,177   & valid $H_V$ and $H_R$                        & 69\,188 \\
$H_i$   & 80\,196   & & \\
$H_z$   & 68\,622   & &  \\
$H_V$   & 72\,785   & &  \\
$H_R$   & 71\,413   & &  \\
\hline
\end{tabular}
\end{table}

\begin{figure}
    \centering
\includegraphics[width=\hsize]{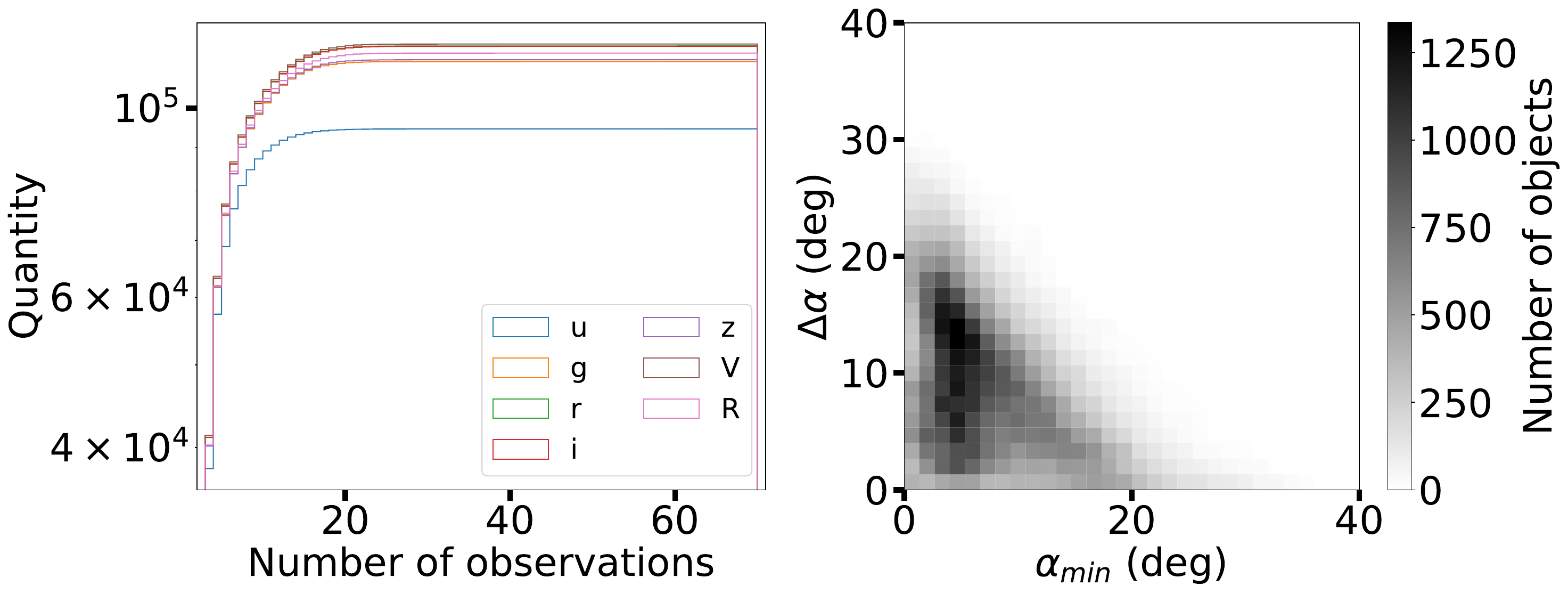}
\caption{Observational circumstances of the data used in this work. Left panel: Cumulative distribution of the number of observations per filter for the objects in this work. Right panel: Minimum $\alpha$ vs. span in $\alpha$ covered by the objects in this work.}\label{fig:dists}%
\end{figure}

Most objects have less than ten observations, with a median of four per filter. Noteworthy, over 100 objects have more than 20 observations, and one object was observed 70 times in the g, r, and i filters (Fig. \ref{fig:dists}, left panel). Most of the objects were observed with $\alpha_{min}<10$ deg with a median span, $\Delta\alpha$, of 8.3 deg (Fig. \ref{fig:dists}, right panel). The vast majority of objects are main belt asteroids (Fig. \ref{fig:orbelem}), but there are objects with $a>100$ AU at high eccentricities to fulfill the criteria set above to compute the PC.

\begin{figure}
\centering
\includegraphics[width=4.4cm]{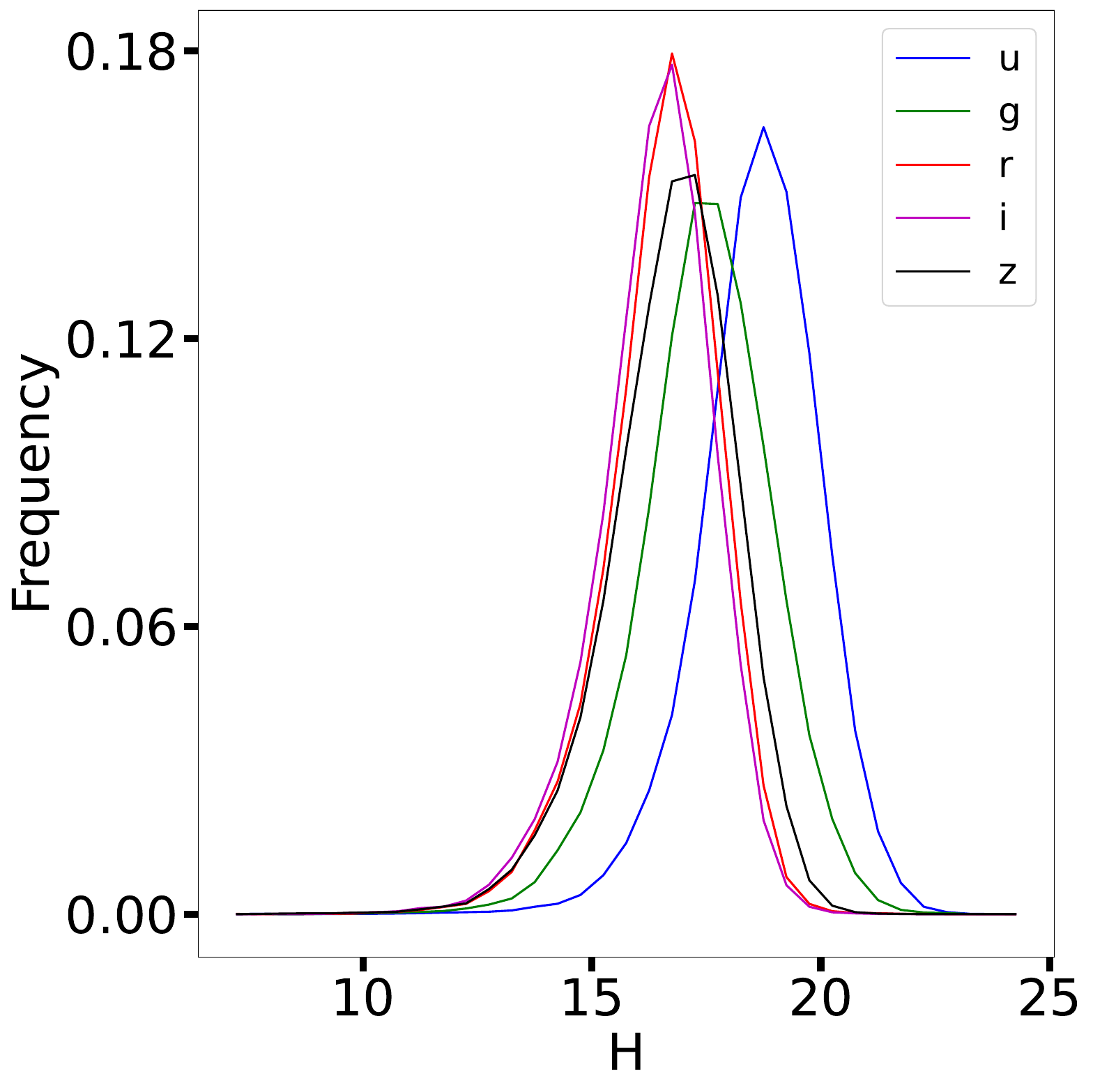}
\includegraphics[width=4.4cm]{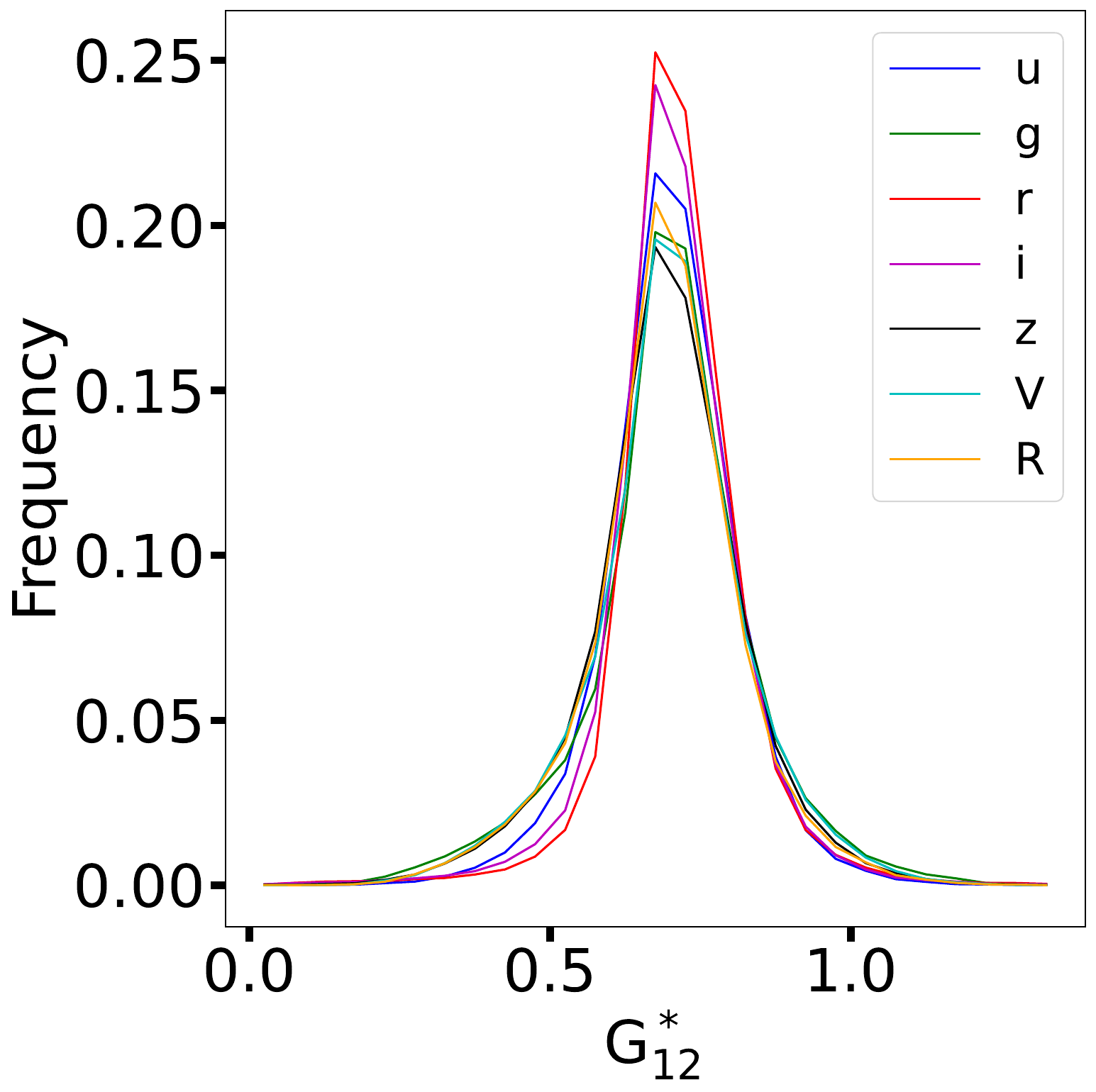}
\caption{Distributions of computed quantities. Distributions of $H$ (left panel) and $G$ (right panel). $u$ is shown in the blue line, $g$ in green, $r$ in red, $i$ in purple, and $z$ in black.}\label{fig:figdists}%
\end{figure}

The distributions of absolute magnitudes and $G$ are shown in Fig. \ref{fig:figdists}.
As already seen in AC22, the $H_u$ distribution peaks at about two magnitudes fainter than those in the other filters (compatible with the solar $(u-g)=1.43$). The $G$ distributions all peak at about 0.6 with a FWHM of about 0.2. There are no apparent changes with wavelength other than a slight width increase towards bluer wavelengths.

\begin{figure}
\centering
\includegraphics[width=4.4cm]{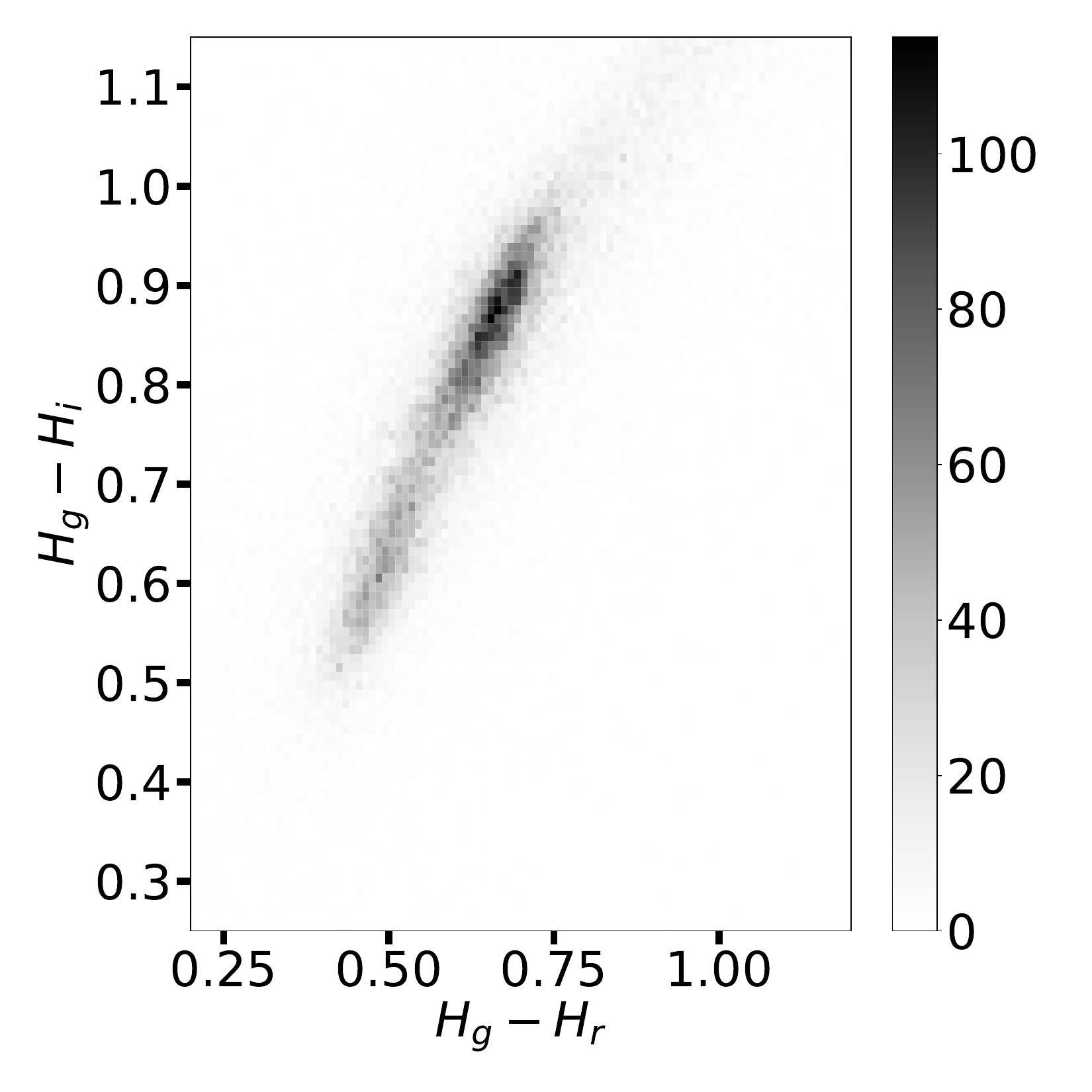}
\includegraphics[width=4.4cm]{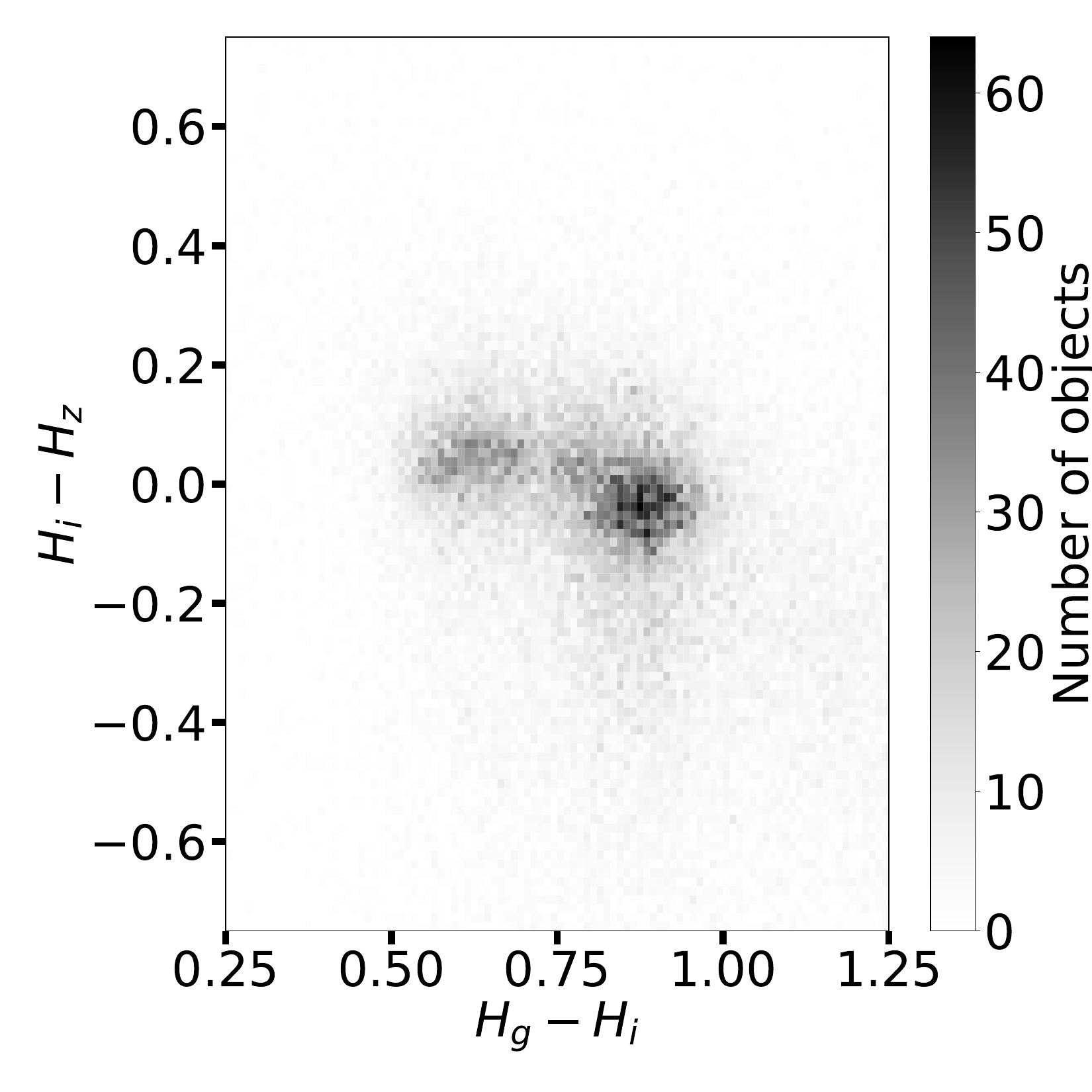}
\caption{Color - color diagrams in the form of {density plots}. The Left panel shows the $H_g-H_r$ vs. $H_g-H_i$, highlighting the linear part of the spectrum, while the right panel shows the $H_g-H_i$ vs. $H_i-H_z$ does the same for the region that encompasses the 900 nm absorption feature due to silicates.}\label{fig:colors}%
\end{figure}

The computed $H_g-H_i$, $H_g-H_r$, and $H_i-H_z$ are displayed in Fig. \ref{fig:colors} in color-color diagrams using density plots to simplify the figures.
Overall, the distributions follow the same trend as in previous works \citep[for example][]{alcan2022,colazo2022,sergeyev2023}. It is worth mentioning that several objects seem to escape from the usual phase space {because they have large uncertainties in their colors} (seen as faint clouds in Fig. \ref{fig:colors}), and care must be taken when using this database.

\subsection{Spectral slope}\label{sec:coloring}
\cite{alcan2022} established that the method was trustful, and it provided comparable results with other methods used to compute $H$ (for example, \citealt{oszki2011pcs} or \citealt{veres2015}). Therefore, we jump straight into the central issue of this article: {\it Phase coloring}.

\begin{figure}
\centering
\includegraphics[width=4.4cm]{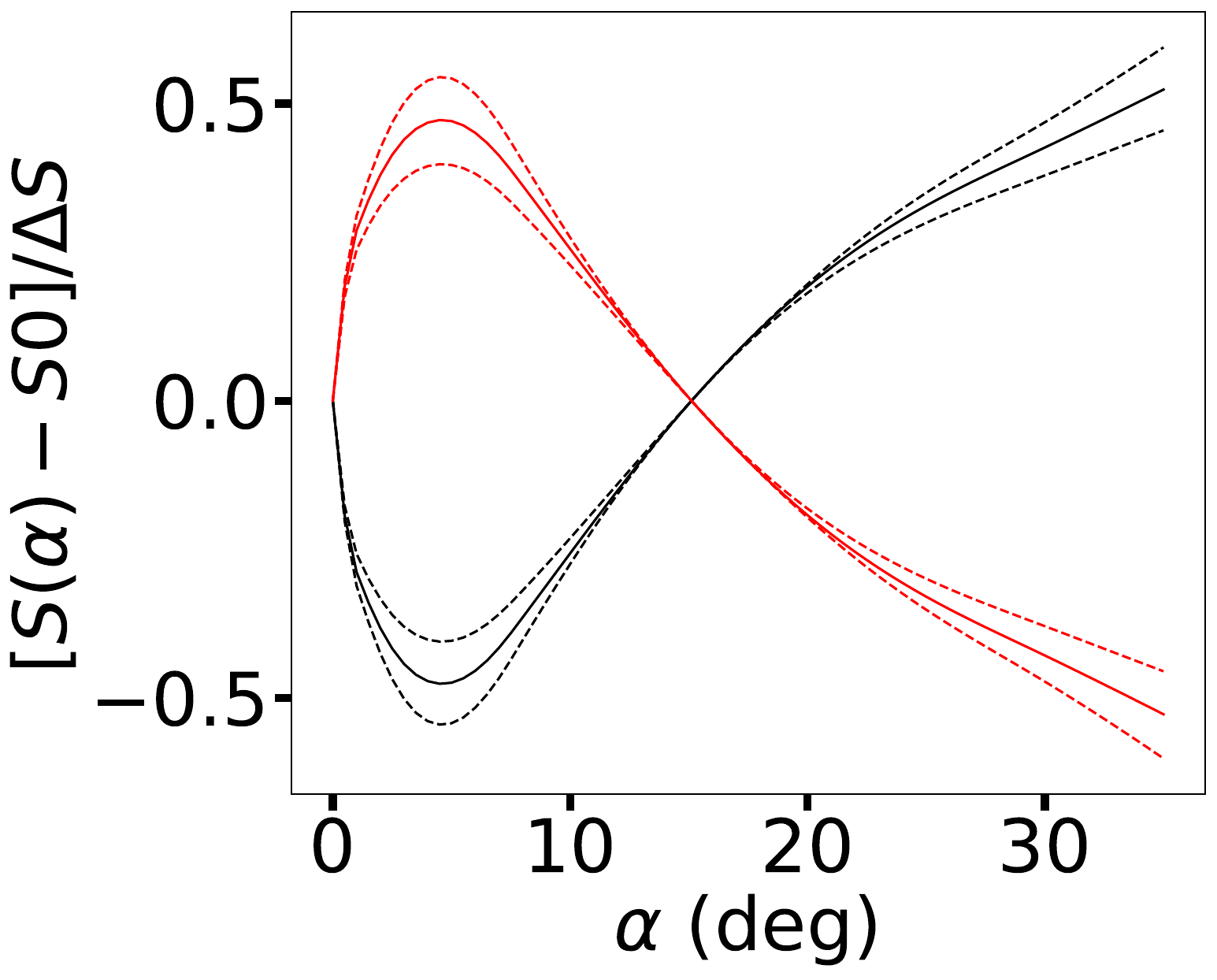}
\includegraphics[width=4.4cm]{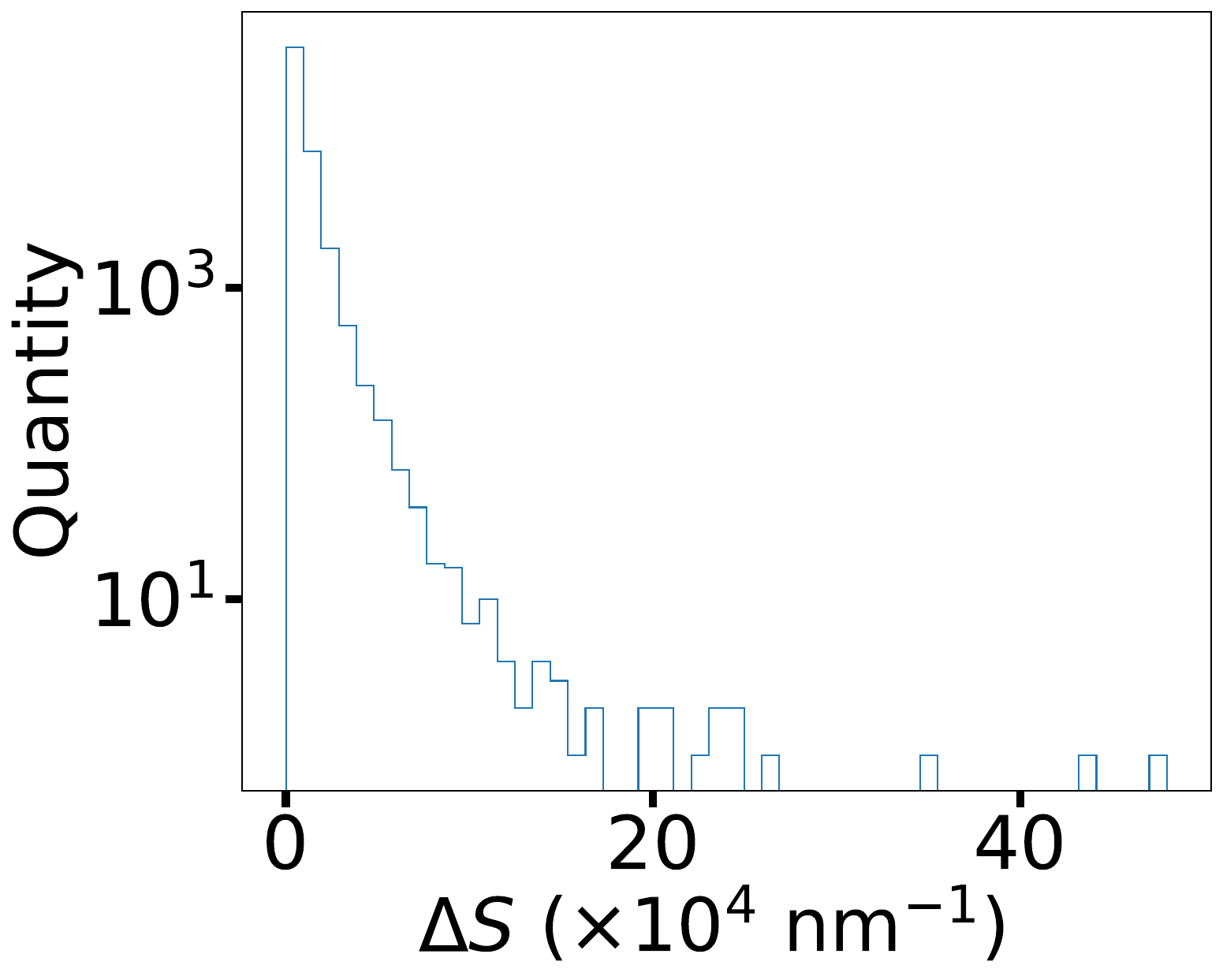}
\caption{The two behaviors of the spectral slope. Left: Change of spectral slope relative to the slope at $\alpha=0$ deg and normalized to its maximum amplitude to increase contrast. The continuous line shows the average value of all objects with that behavior, while the dashed lines indicate the maximum and minimum values per $\alpha$ bin. Right: {Distribution of $\Delta S$.} }\label{fig:slope0}%
\end{figure}

We compute first the spectral slope (slope for short) at different $\alpha$ for all objects {with $H_u$, $H_g$, $H_r$, and $H_i$ using a simple linear fit to the computed fluxes (see Eq. \ref{mag_flux}). $H_z$ was excluded from the fit because it falls on the silicate absorption feature in some objects, departing from the linear part of the spectrum.

The fluxes are computed using}
\begin{equation}\label{mag_flux}
F_{\lambda}(\alpha) = 10^{-0.4[((M_{\lambda} - M_r)(\alpha)-(M_{\lambda} - M_r)_{\odot}]},
\end{equation}
where the solar colors were taken from the {SDSS\footnote{https://live-sdss4org-dr14.pantheonsite.io/algorithms/ugrizVegaSun/}.}
{Note that in Eq. \ref{mag_flux}, it appears the object's color as an explicit function of the phase angle. The color was computed using
the nominal values of $H$s and $G$s, feeding them into Eq. \ref{eq:2}. The computation was done in steps of 0.5 deg between $\alpha\in[0,35]$ deg.} Figure \ref{fig:slope0} shows the results for 42\,168 objects. The y-axis shows a normalized spectral slope to ease comparison. This slope is computed as
\begin{equation}\label{eq:6}
    {\rm Norm. slope} =\frac{ S(\alpha) - S0}{\Delta S},
\end{equation}
where $\Delta S=max[S(\alpha) - S0] - min[S(\alpha) - S0]$, and $S0=S(\alpha=0 {\rm~deg})$.
It is immediately clear that there are two types of curves: (i) some slopes decrease with $\alpha$ until a critical value $\alpha_c$, and then revert the behavior {(the BR group)}; (ii) other curves increase the slopes until a critical value, and then starts decreasing {(the RB group)}. $\alpha_c$ seems to lie between 4.5 and 5 degrees in all cases. {Behavior BR occurs for 22\,858 objects and RB for 19\,310}. Note that there are no flat curves. {Therefore, an object falls either on the BR or the RB group}. In practice, if the normalization factor in Eq. \ref{eq:6} is removed, some curves will appear flat due to the small amplitude of $\Delta S$. To highlight the maximum amount of change in $S(\alpha)$, we plot a histogram with the distribution of $\Delta S$, as defined above, in Fig. \ref{fig:slope0} (left). It is clear that most objects do not display a radical change of $S(\alpha)$; the median value is $0.47\times10^{-4}$ nm$^{-1}$, within usual uncertainties in the measurements of spectral slopes. Nevertheless, it is important to note that there are almost 300 objects with $\Delta S>5\times10^{-4}$ nm$^{-1}$, which is well above typical uncertainties. In any case, we would like to stress that the results displayed in Fig. \ref{fig:slope0} should be taken as a {\it toy model} representation of the results, as it does not include the full information in the probability distributions, although it is useful to glimpse the existence of different phase-coloring behaviors.

\begin{figure}
\centering
\includegraphics[width=6.5cm]{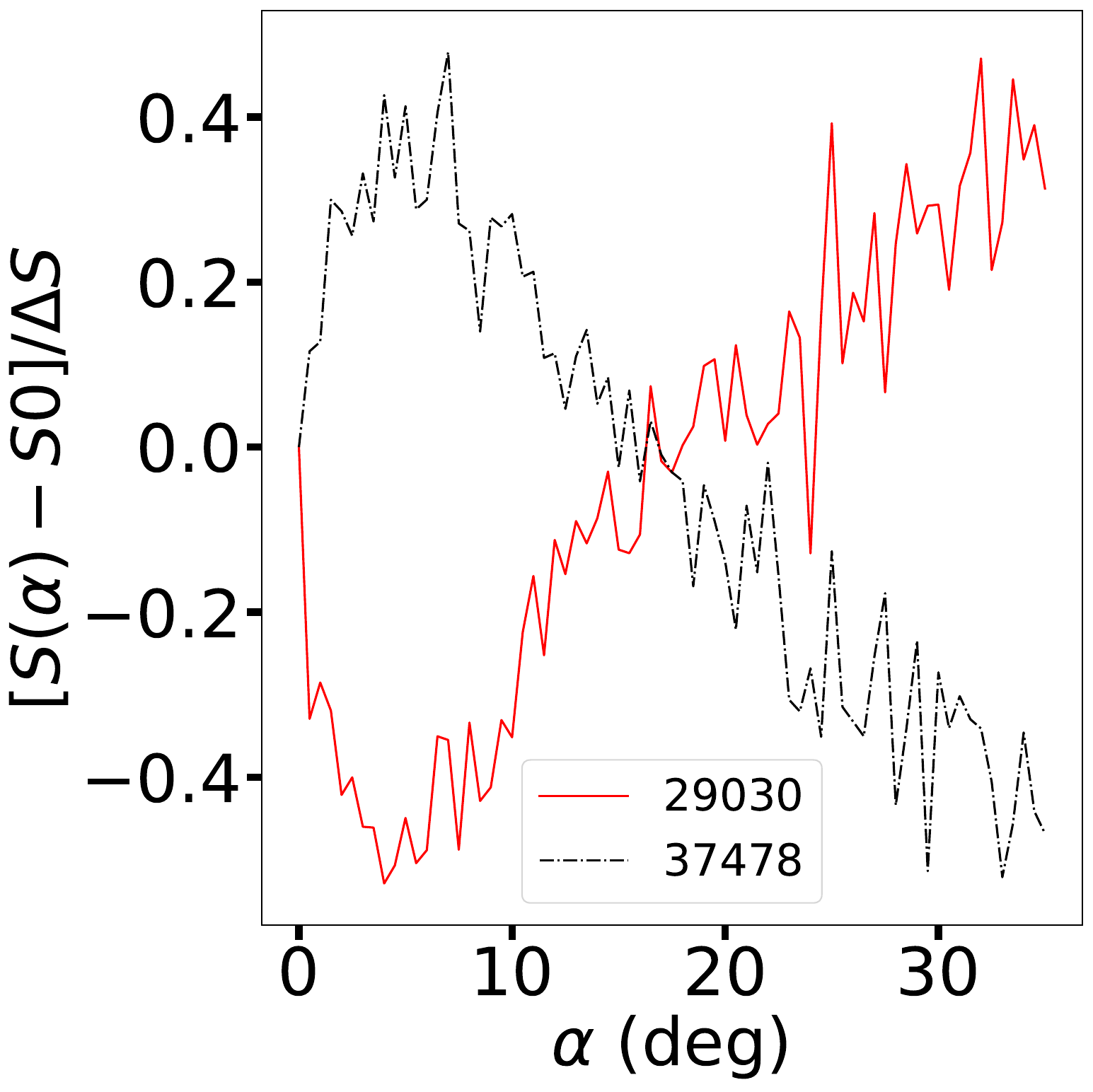}
\caption{Change of spectral slope relative to the slope at $\alpha=0$ deg and normalized to its maximum amplitude to increase contrast using the full $P(H_{\lambda})$. In the continuous red line, we show the curve for 29030, {whose colors make it possibly an S-complex asteroid, see below}, while in the dot-dashed black line, we show the curve for 37478, {a possible C-complex asteroid, according to its colors.}}\label{fig:slope1}%
\end{figure}

The above results were obtained using the modeled data, which does not account for realistic observations. To add more ``reality'', we did a second experiment using the full probability distribution of each absolute magnitude, extracting 1\,000 different random values from each $H$ and its corresponding $G$, obtaining a distribution of slopes per object and $\alpha$. In this case, Fig. \ref{fig:slope0} cannot be reproduced because of the noise in the solutions. Nevertheless, inspecting some of the resulting curves visually, it is possible to detect that the behavior is discernible in high-quality data (Fig. \ref{fig:slope1}).

\begin{figure}
\centering
\includegraphics[width=4.4cm]{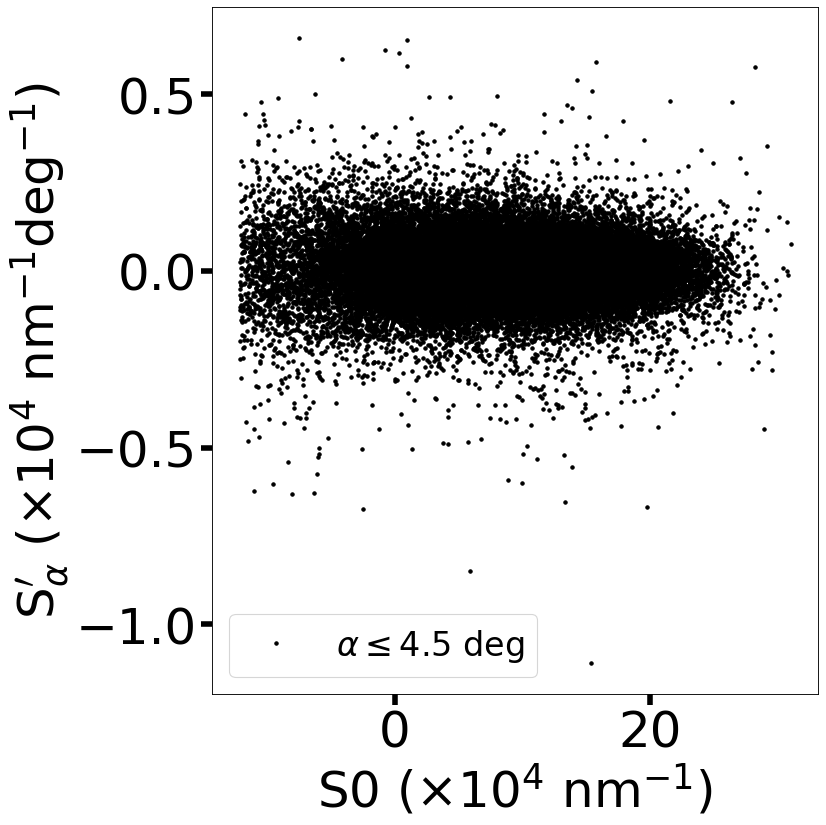}
\includegraphics[width=4.4cm]{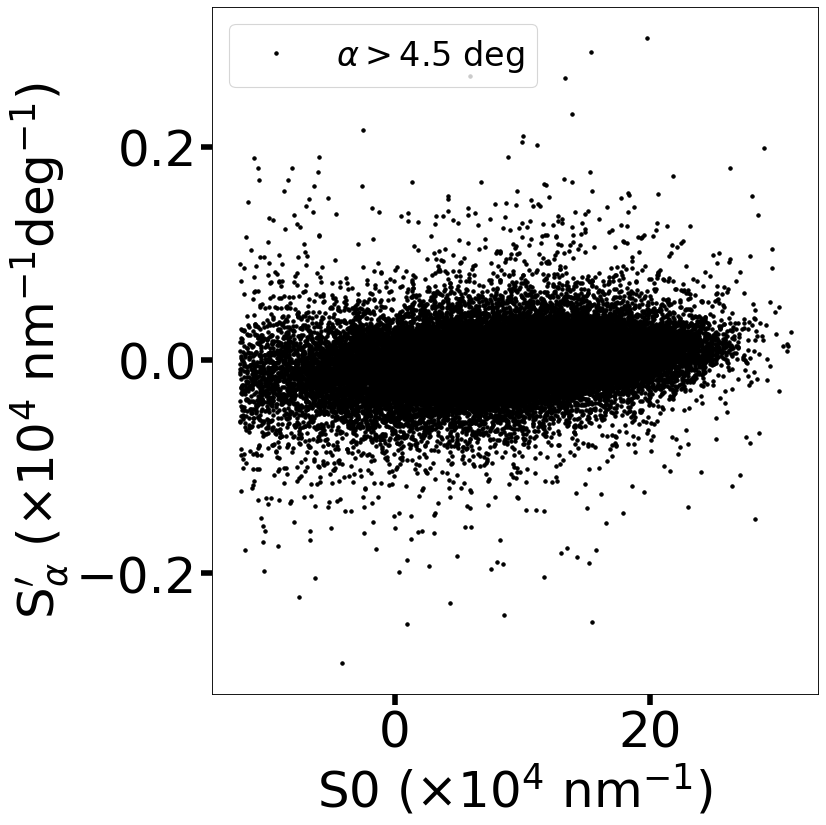}
\caption{Change of spectral slope ($S'_{\alpha}$) with respect to the spectral slope at opposition, $S0$. Left, $S'_{\alpha}$ computed for $\alpha\leq4.5$ deg, Right, $S'_{\alpha}$ computed for $\alpha>4.5$ deg.}\label{fig:szerovss_alpha}%
\end{figure}

{Given the results shown in Fig. \ref{fig:slope0} and \ref{fig:slope1}, we decided to separate the sample into two regimes and compute the change of slope ($S'_{\alpha}$) with phase angle between 0 and 4.5 deg, and between 5.0 and 35 deg. $S'_\alpha$ is computed by a linear fit to $S(\alpha)$ vs. $\alpha$ in both cases. Both cases' results were clipped (at $3\sigma$) to remove outliers.} The results are shown in Fig. \ref{fig:szerovss_alpha}. The left panel shows, {\it a priori}, no relation between $S'_{\alpha}$ and $S0$. The Spearman test shows that there is a weak negative anti-correlation with a $r_s = -0.01$ and a p-value of 0.0026, in a way resembling (though much weaker) that detected for asteroids at small phase angles by \cite{alvcan2022}, using a different photometric model. The right panel of the figure shows that above 4.5 deg, the general tendency is for objects to become redder when they are intrinsically redder in opposition.

\begin{figure}
\centering
\includegraphics[width=4.4cm]{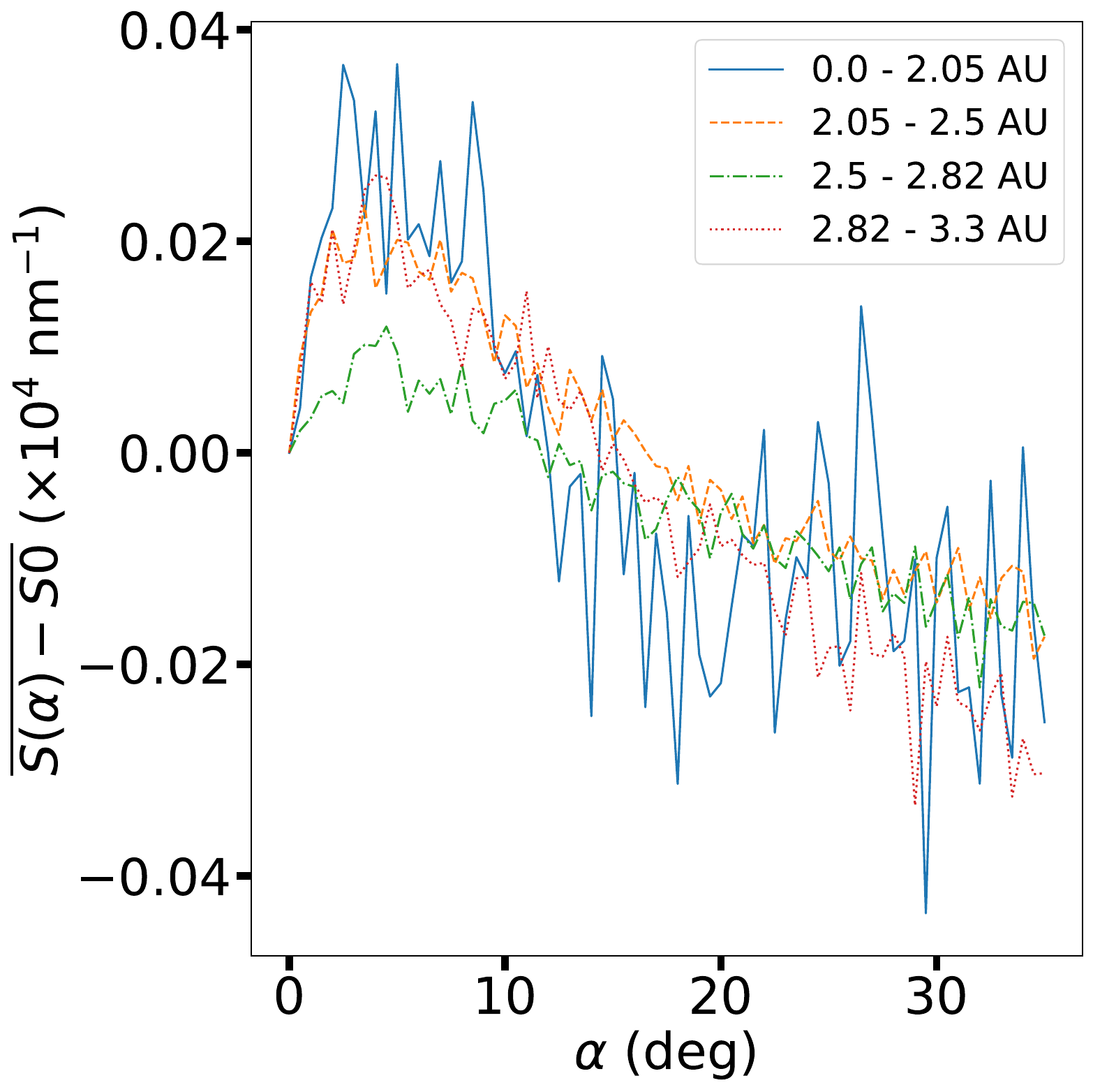}
\includegraphics[width=4.4cm]{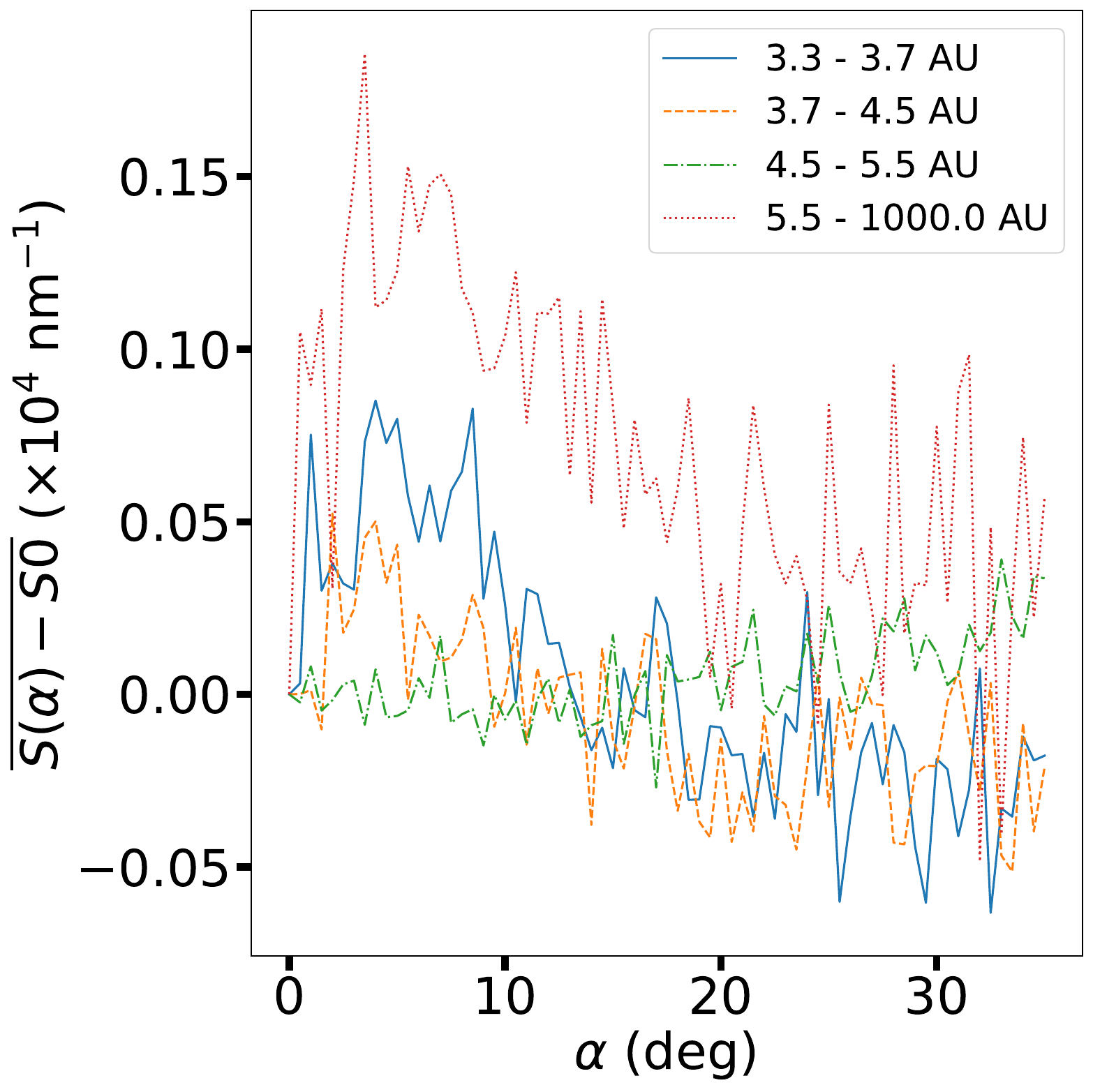}
\caption{Average relative spectral slope in bins of semi-major axis. The results for each bin are marked within the figures.}\label{fig:avslopes}%
\end{figure}

In the next step, we checked for the coloring behavior in different bins of semi-major axis. The data was split into eight regions: (i) $a\leq 2.05$ AU, (ii) $2.05\le a\leq 2.5$ AU, (iii) $2.5\le a\leq 2.82$ AU, (iv) $2.82\le a\leq 3.3$ AU, (v) $3.3\le a\leq 3.7$ AU, (vi) $3.7\le a\leq 4.5$ AU, (vii) $4.5\le a\leq 5.5$ AU, and (viii) $a>5.5$ AU. We computed the average $S(\alpha)-S0$ value within each region per phase angle bin. The results are shown in Fig. \ref{fig:avslopes}.
The figure clearly shows that the behavior of the average relative slope is not monotonic, neither with phase angle nor in the semi-major axis. Most curves show a reddening of the slopes up to 4.5 deg, then the relative slope decreases. The only ``odd'' behavior occurs in the semi-major axis occupied by the Jovian Trojans, {$4.4-5.5$ AU}, where there is bluing and then reddening of the relative slopes. {We checked that the distribution of $\alpha$ is similar in all semi-major axis bins to avoid introducing undesired biases.}

The uncertainties of the quantities were not considered to avoid blurring the results too much. Note, however, that uncertainties were computed for the realistic slope as the standard deviation of the distributions and are available for the reader in the OSF folder.

\subsection{Colors and phase angle}\label{sec:colors}

\begin{figure}
\centering
 \includegraphics[width=4.4cm]{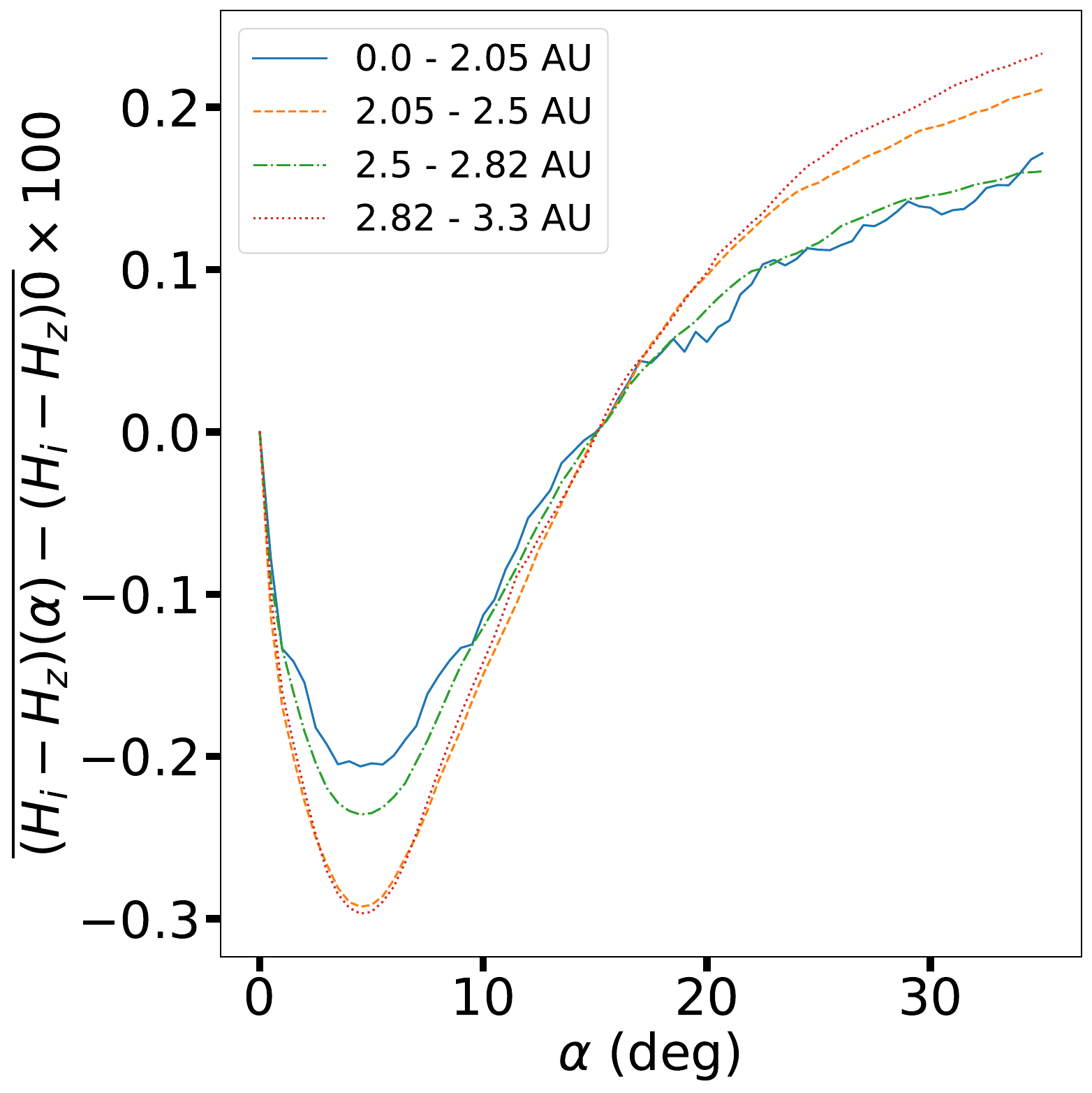}
 \includegraphics[width=4.4cm]{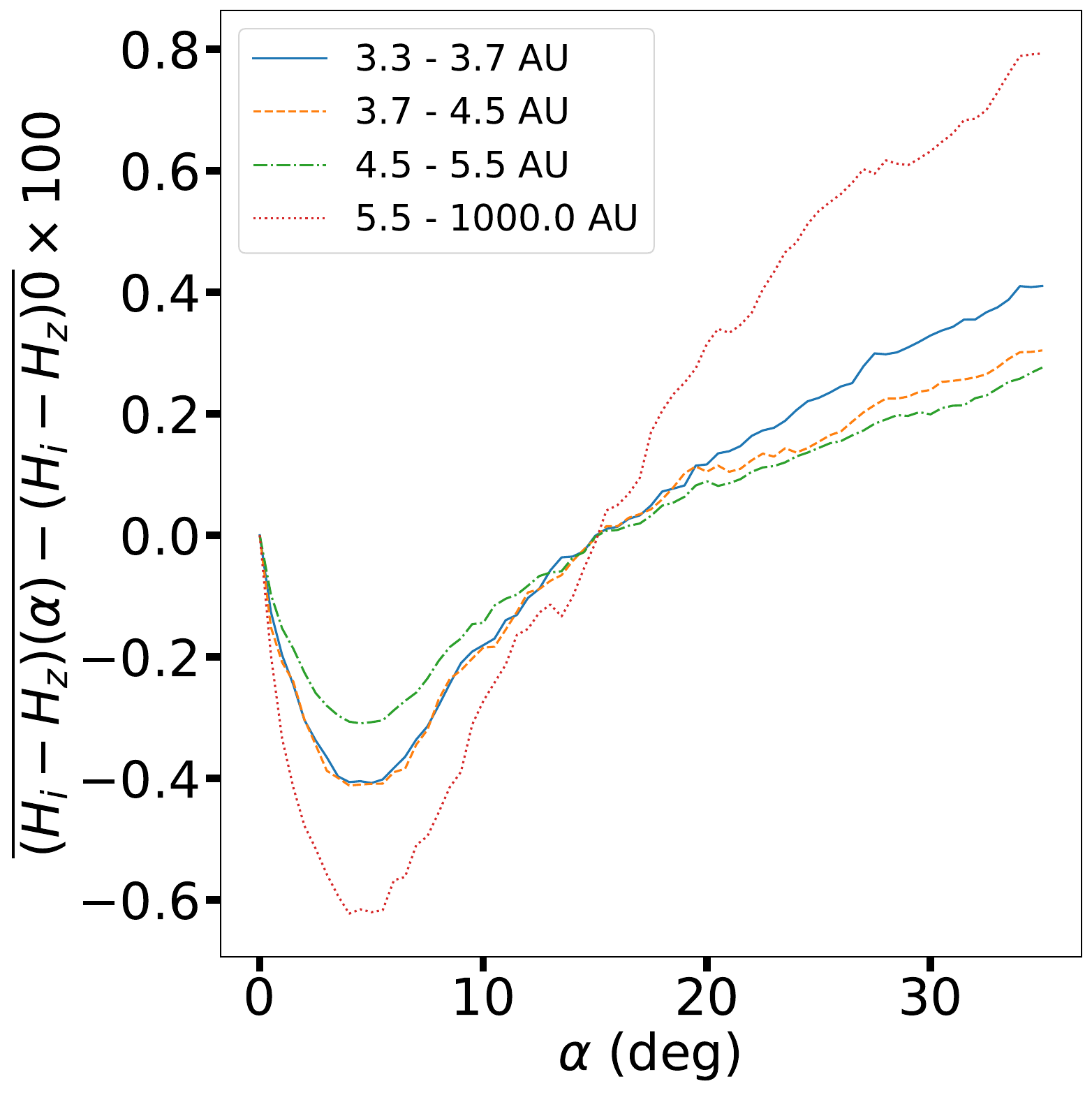}
\caption{Average relative $H_i-H_z$ in bins of semi-major axis. The results for each bin are marked within the figures.}\label{fig:aviz}%
\end{figure}

\begin{figure}
\centering
\includegraphics[width=4.4cm]{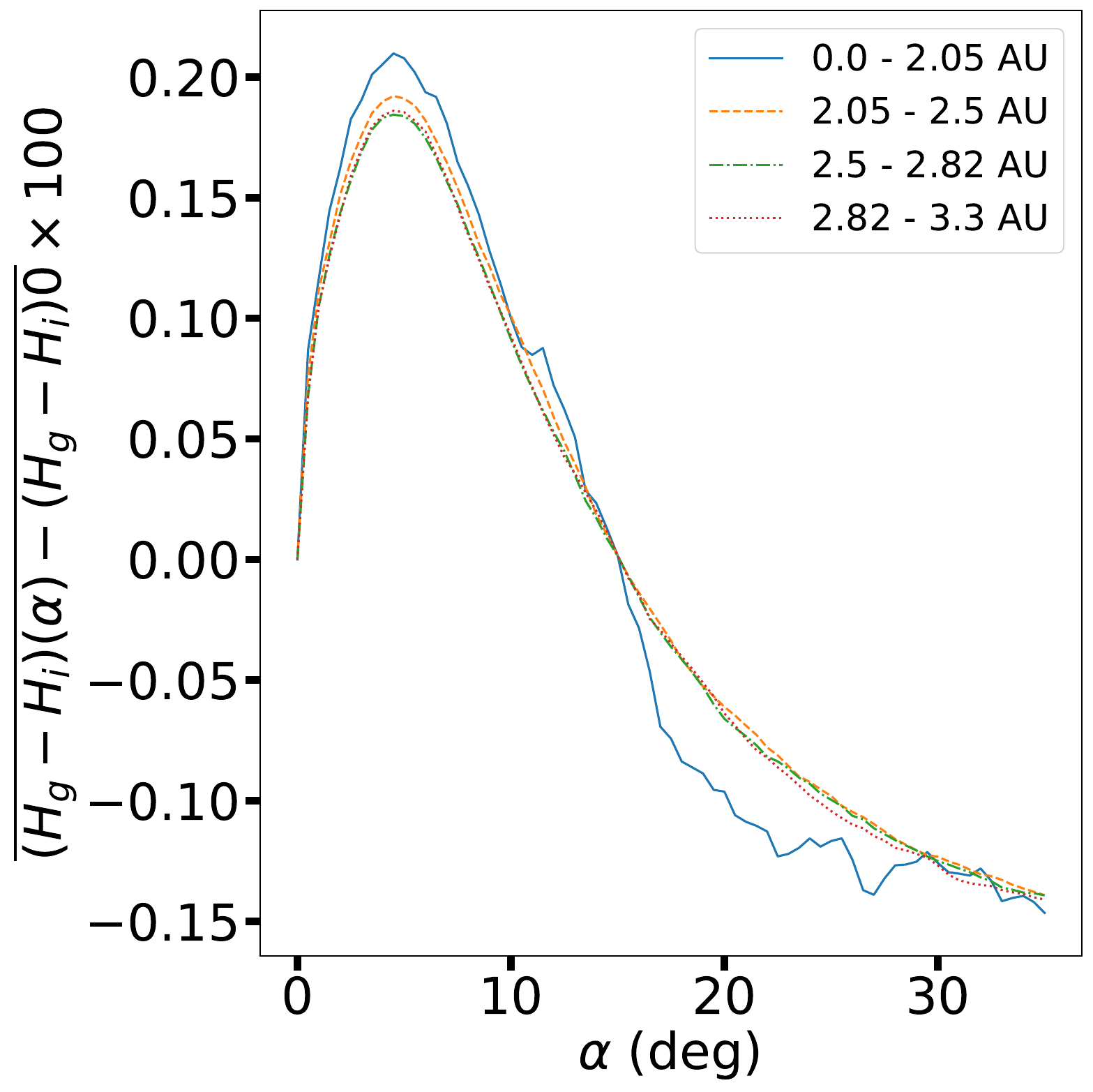}
\includegraphics[width=4.4cm]{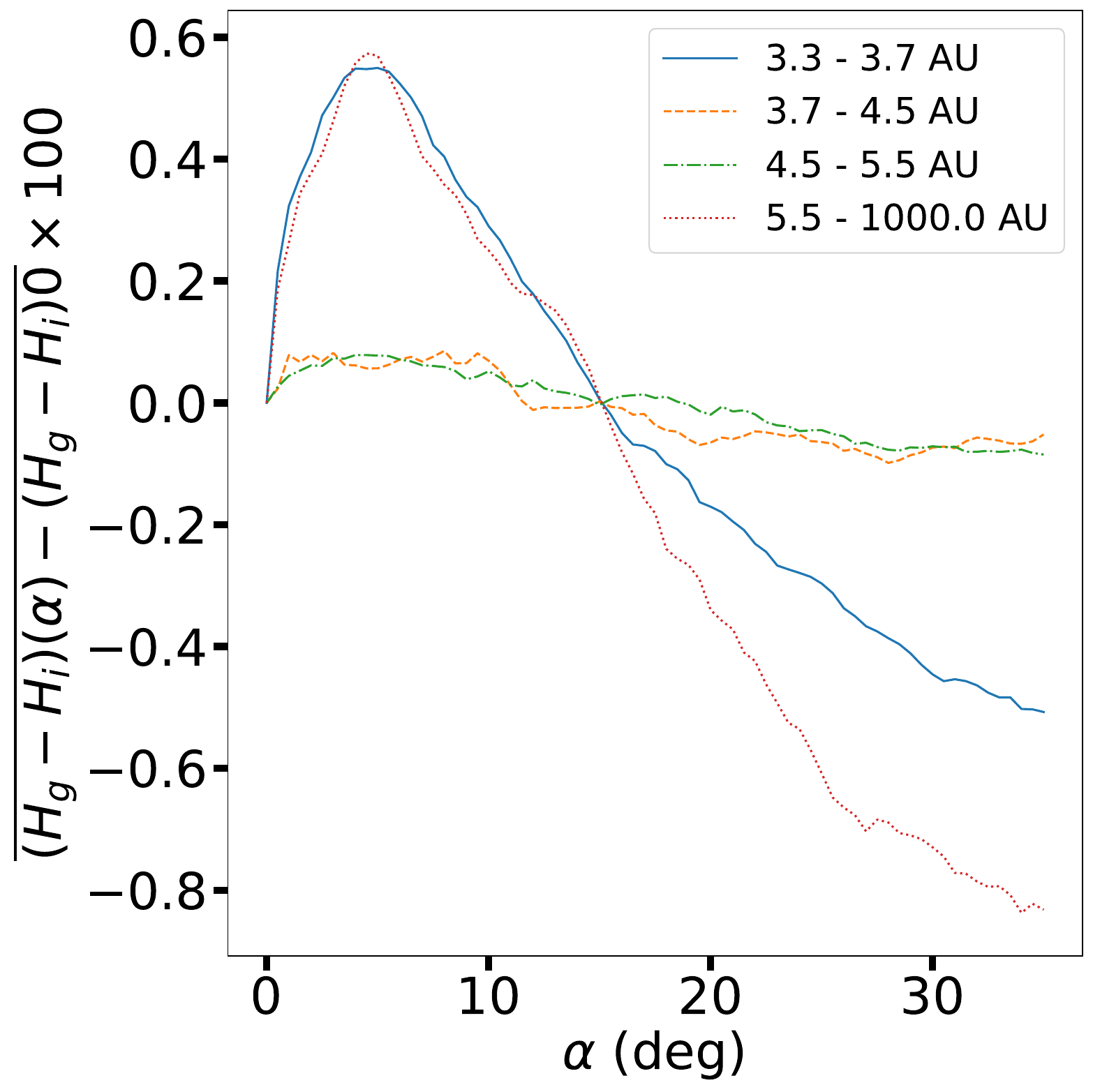}
\caption{Average relative $H_g-H_i$ in bins of semi-major axis. The results for each bin are marked within the figures.}\label{fig:avgi}%
\end{figure}

Not only the evolution of the slope with $\alpha$ is interesting. We also studied the change of the colors with the phase angle. Following a similar procedure as above to compute the more realistic slopes, we use the average relative color in the same semi-major axis bins as above and show the results for the $H_i-H_z$ in Fig. \ref{fig:aviz} and for $H_g-H_i$ in Fig. \ref{fig:avgi}
The behavior of $H_i-H_z$ indicates that, on average, the color decreases until about 5 deg in all semi-major axis bins, and it increases, getting redder with increasing $\alpha$. Note that objects of all taxa are included in the averaging.

On the other hand, $H_g-H_i$ seems to follow the curves of {the average relative slope (Fig. \ref{fig:avslopes})}, which is natural since this region is less affected by the 900 nm absorption band. Overall, the color seems to increase until the critical angle, and the color starts decreasing after $\alpha_c$. The same trends are seen in all semi-major axis bins, while the Hilda and Trojan regions have the less dramatic changes.

As mentioned, the uncertainties of the quantities were not included to avoid blurring the results. {Nevertheless, the uncertainties were computed as the interval between the 16$^{\rm th}$ and the 84$^{\rm th}$ of the convolution of the PDFs of the colors}, as explained in Sect. 3.2 of \cite{alvcan2022}. These uncertainties are available for the reader in the OSF folder.

\subsection{The impact on the taxonomy classes}\label{sec:taxa}
{The previous sections established that spectral slopes and colors change with phase angle. The next step is to check whether these changes are sufficient to change the taxon of an object. To do so, a simple experiment was performed using the four taxa defined by \cite{colazo2022}, i.e., C, X, S, and V complexes. We crudely separated the $H_g-H_i$ vs. $H_i-H_z$ into the four regions each taxon covers (Fig. \ref{fig:taxa} provides a schematic view). The C-complex is located in $H_g-H_i\in(0.47,0.727)\wedge H_i-H_z\in(-0.128,0.25)$, the X-complex in $H_g-H_i\in(0.727,1.05)\wedge H_i-H_z\in(1.129[H_g-H_i]-0.926,0.25)$, the S-complex in in $H_g-H_i\in(0.727,1.05)\wedge H_i-H_z\in(-0.128,1.129[H_g-H_i]-0.926)$. Finally, the V-complex is in $H_g-H_i\in(0.727,1.05)\wedge H_i-H_z\in(-0.478,-0.128)$.
\begin{figure}
\centering
\includegraphics[width=6.5cm]{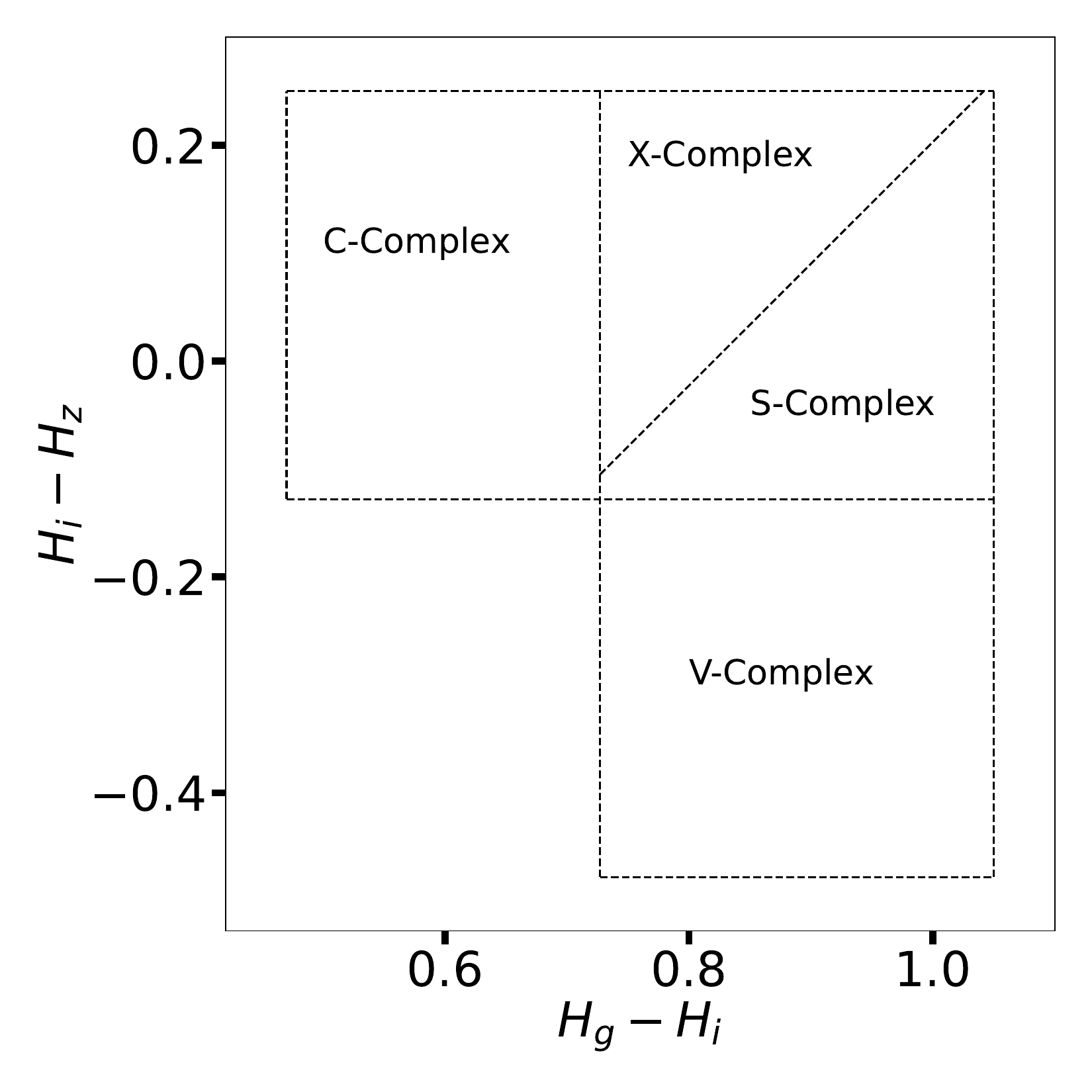}
\caption{Schematic view of the boxes considered for each taxon.}\label{fig:taxa}%
\end{figure}

We computed the number of objects in each box at $\alpha=0$ deg, $N_0$, the number $N_c$ at $\alpha_c$, and the number of objects, $N_m$, at the maximum phase angle ($\alpha_{max}$). Notice that, according to Figs. \ref{fig:aviz} and \ref{fig:avgi}, the maximum differences in color should happen between $\alpha_c$ and $\alpha_{max}$. The results are shown in Table \ref{table:3}, where the differences are expressed in per mille.
\begin{table}
\caption{Relative change of taxa}
\label{table:3}
\centering
\begin{tabular}{c c | c c}
\hline\hline
Complex  & $N_0$ & $\frac{N_c-N_0}{N_0}$ (\textperthousand) & $\frac{N_m-N_c}{N_c}$ (\textperthousand) \\
\hline
C   & 7\,364   &  15 & -31 \\
X   & 4\,574   &  13 & -31 \\
S   & 8\,156   &   7 & -23 \\
V   & 5\,817   & -15 &  25 \\
\hline
\end{tabular}
\end{table}

As expected, the largest changes happen between $N_c$ and $N_m$. Curiously, the V-complex is the only one that {loses objects from opposition to $\alpha_c$ to increment their numbers from $\alpha_c$ to $\alpha_{max}$}. In any case, this is a rough estimate that provides minimum numbers because the space covered by each box is larger than the actual taxon should occupy as well as we did not follow any particular object but just counted the number of objects at each step; therefore, it is likely that between steps more objects moved in or out of each region. (See also Appendix \ref{aapA}.)}

\subsection{On the use of weighted averages}
Weighted arithmetic averages are commonly used to compute colors (and other quantities) of Solar System objects \citep[for example][]{mboss2012,sergeyev2023}. The idea behind this approach is to prioritize data with high precision, the usual weight being the inverse of the uncertainty (or square uncertainty). This approach is valid if, and only if, all data are compatible. In this case, by compatible, we mean if all measurements were taken at the same phase angle, {distances to the Sun and Earth, and rotational phase.} Otherwise, some systematic may be introduced as objects tend to get brighter closer to the opposition, presumably increasing the precision of the measurements.

\begin{figure}
\centering
\includegraphics[width=6.5cm]{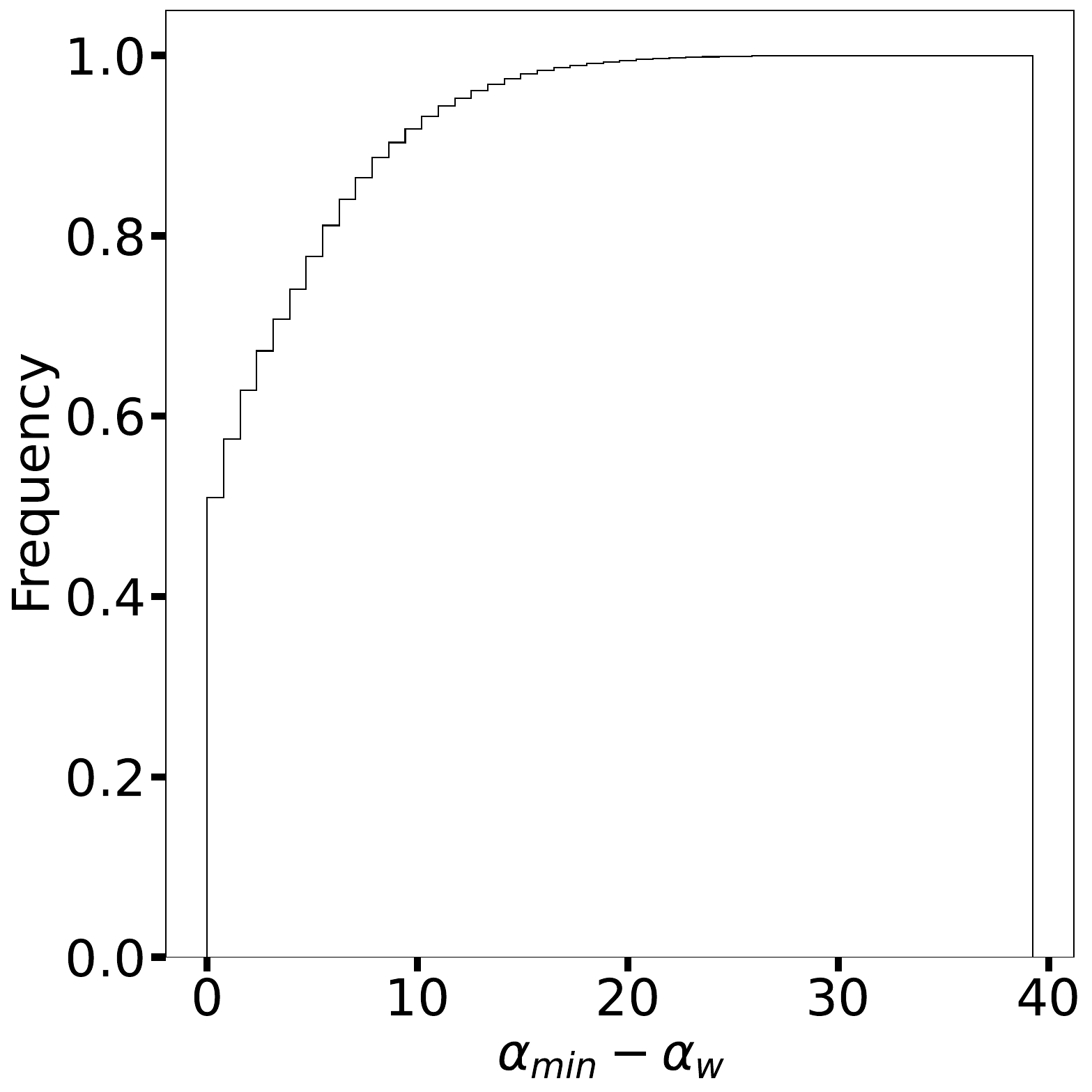}
\caption{Cumulative distribution of $\alpha_{min}-\alpha_w$ (see text for a description of both angles).}\label{fig:aw-am}%
\end{figure}

We looked into S21's data and used the same criteria as mentioned above to select the data, but in this case, computing the weighted average in $u-g$, $g-r$, $r-i$, and $i-z$ {and then comparing these averaged value with the individual colors at each value of $\alpha$ and saved the phase angle (called $\alpha_w$) where they were closer.} Then, we compared the value of $\alpha_w$ with the minimum phase angle for each object. {The results displayed in Fig. \ref{fig:aw-am} correspond to the $g-r$ color measurements and are similar in the other colors.}
In the figure, it is clear that using a weighting scheme biases the colors towards low-phase angles (i.e., closer to opposition): 50\% of the average colors coincide with the minimum $\alpha$ in the phase curve within 1 deg, while 3/4 of the sample are within 5 deg of the minimum phase angle. Therefore, we recommend care when using averaged values because these values are biased toward low-phase angle measurements carrying more weight. To further complicate the interpretation of results obtained with average colors, \cite{alcan2022} {showed that the average colors and the colors at opposition (i.e, the difference of absolute magnitudes that we call absolute colors)} can have significant differences and that there is no one-to-one correspondence between them.

\section{Conclusions}\label{sect:conclussions}
The terminology {\it Phase reddening} has been used for a while to describe an observational effect: objects observed at higher phase angles tended to have redder colors (or larger spectral slopes \citealt[for instance]{luujewitt1990AJ}). Nevertheless, this was usually seen in a relatively low number of objects due to the inefficiency of traditional observation programs in scheduling repeated observations of the same object in different phase angles. For example, \cite{Nathues2010Icar}, from a sample of over 90 Eunomia objects, only have six objects with two or more observations, including one with a clear bluing: (5785) Fulton.

The results we show in this paper point to different behaviors in different phase angles regimes: In terms of spectral slope, there seems to be a slight anti-correlation between $S'_{\alpha}$ computed for $\alpha\leq4.5$ deg and $S0$, suggesting that intrinsically redder objects tend to get slightly bluer with increasing phase angle, while the opposite is seen for intrinsically bluer objects (see Fig. \ref{fig:szerovss_alpha}), these results are strikingly similar to these of \cite{ayala2018} for TNOs and \cite{alvcan2022} for asteroids. On the other hand, $S_{\alpha}$ for $\alpha>5$ deg shows, on average, that intrinsically redder objects tend to get redder in line with the traditional view of phase coloring.

There also seem to be two behaviors in terms of spectral slopes with respect to $\alpha$ (Figs \ref{fig:slope0}, \ref{fig:slope1}, and \ref{fig:avslopes}): The objects tend to have one behavior in the range $[0,4.5]$ deg, while the opposite in the range $(4.5,35]$ deg. There seems to be a slight predominance of objects that increase the spectral slope and then get bluer{, the RB group}. This behavior happens in almost all semi-major axis bins, but the effect is within error bars in typical spectral slopes measurements (the unit [$10^4$ nm$^{-1}$] is equivalent to [\% (1000~\AA)$^{-1}$]). {We also cross-matched the data with the taxa determined by \cite{colazo2022} to test if the two behaviors could be related to surface composition without finding any evidence.}
{In the Appendix \ref{appB} we discriminate the behavior of the RB and BR groups to analyze their different demeanor.}

Color-wise, there are similar behaviors: first, going in one direction to suddenly change in the opposite. The $H_g-H_i$ correlates with the slope, which is expected because the i filter falls in a region relatively unaffected by absorption features. We find the behavior of the $H_i-H_z$ more interesting because it targets the absorption band at 900 nm due to silicates. In this case, the color goes bluer for $\alpha\in[0,4.5]$ deg, while the average color increases for $\alpha>4.5$ deg. Therefore, the band gets deeper or shallower in objects with an absorption feature just by phase coloring. Joining the results for both colors, $H_g-H_i$ and $H_i-H_z$, there seem to be objects to get more concave for low-$\alpha$ and less concave for higher phase angle. Once again, we stress the orders of magnitude of the effect, which is at most about one-hundredth of magnitude.

{One interesting point to consider is that $\alpha_c$ coincides roughly with the onset of the non-linear part of the PC, where lies the opposition effect (OE). The OE happens due to a combination of shadow-hiding \citep{hapke1963JGR} and coherent back-scattering \citep{muino1989OE} and produces a net increase in the brightness of the object when observed close to opposition ($\alpha\approx0$ deg). The effect starts to appear below 5 to 6 deg \citep{belskayashev2000}, but it can start as close to the opposition as below 1 deg \citep[for example, in trans-Neptunian objects, see,][]{verbi2022}. The apparent coincidence is very suggestive, and it deserves further analysis beyond this work's scope.} {Perhaps is it here were the physical explanation for the phenomena described in this work lie: which scattering mechanism dominate and how does it behave with the wavelength. Unfortunately, we cannot provide a satisfactory answer yet.}

The effects we report in this paper are unlikely to change how we see the distribution of small bodies' taxa in the Solar System {\citep[for example][]{mothediniz2003Icar,demeo2013Icar,marsset2022AJ,sergeyev2023}}, at least not for the core of the groups, but it does add noise to the frontiers between taxa {where some objects do change taxa. The numbers are a priory small, in the range of a few dozen per thousand. Nevertheless,} with the arrival of massive multi-wavelength photometric surveys, this may change, as the precision in the measurements may reach the level of those shown in this work or close to it. On the other hand, with the large numbers, we can now statistically study the phase-related behavior of many objects, making possible analysis that could not have been done before.

\begin{acknowledgements}
I am grateful to an anonymous reviewer who provided useful comments that helped improve this draft.
AAC acknowledges financial support from the Severo Ochoa grant CEX2021-001131-S funded by MCIN/AEI/10.13039/501100011033.

Funding for the creation and distribution of the SDSS Archive has been provided by the Alfred P. Sloan Foundation, the Participating Institutions, the National Aeronautics and Space Administration, the National Science Foundation, the U.S. Department of Energy, the Japanese Monbukagakusho, and the Max Planck Society. The SDSS Web site is http://www.sdss.org/.

The SDSS is managed by the Astrophysical Research Consortium (ARC) for the Participating Institutions. The Participating Institutions are The University of Chicago, Fermilab, the Institute for Advanced Study, the Japan Participation Group, The Johns Hopkins University, the Korean Scientist Group, Los Alamos National Laboratory, the Max-Planck-Institute for Astronomy (MPIA), the Max-Planck-Institute for Astrophysics (MPA), New Mexico State University, University of Pittsburgh, University of Portsmouth, Princeton University, the United States Naval Observatory, and the University of Washington.

I am grateful to the anonymous reviewer of AC22 who inspired the modification of the algorithm shown in Sect. \ref{sec:analysis}.

This work used https://www.python.org/, https://numpy.org \citep{numpy}, https://www.scipy.org/, and Matplotlib \citep{hunte2007}.

All figures in this work can be reproduced using the following Colab Notebook {https://colab.research.google.com/drive/1-UhSr44YuOAgwMfr-YoWsGztx-QYqgxX?usp=sharing}.

\end{acknowledgements}

\begin{appendix}
\section{Evolution of taxa with $\alpha$}\label{aapA}
\begin{figure*}
    
\centering
 \includegraphics[width=4.4cm]{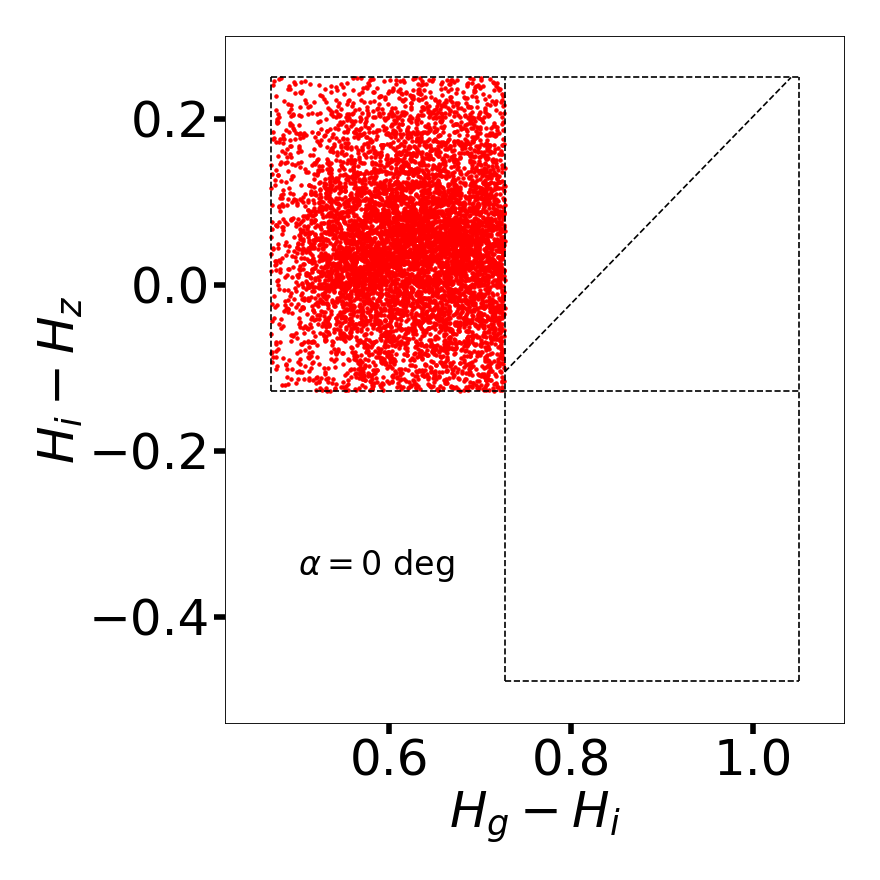}
 \includegraphics[width=4.4cm]{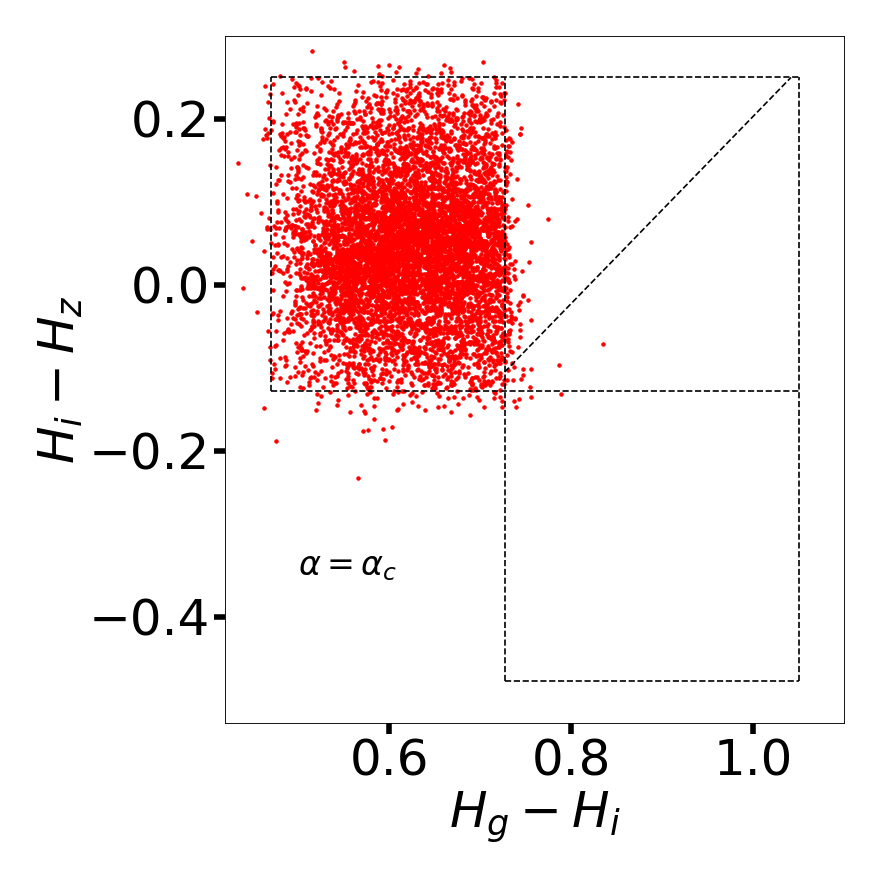}
 \includegraphics[width=4.4cm]{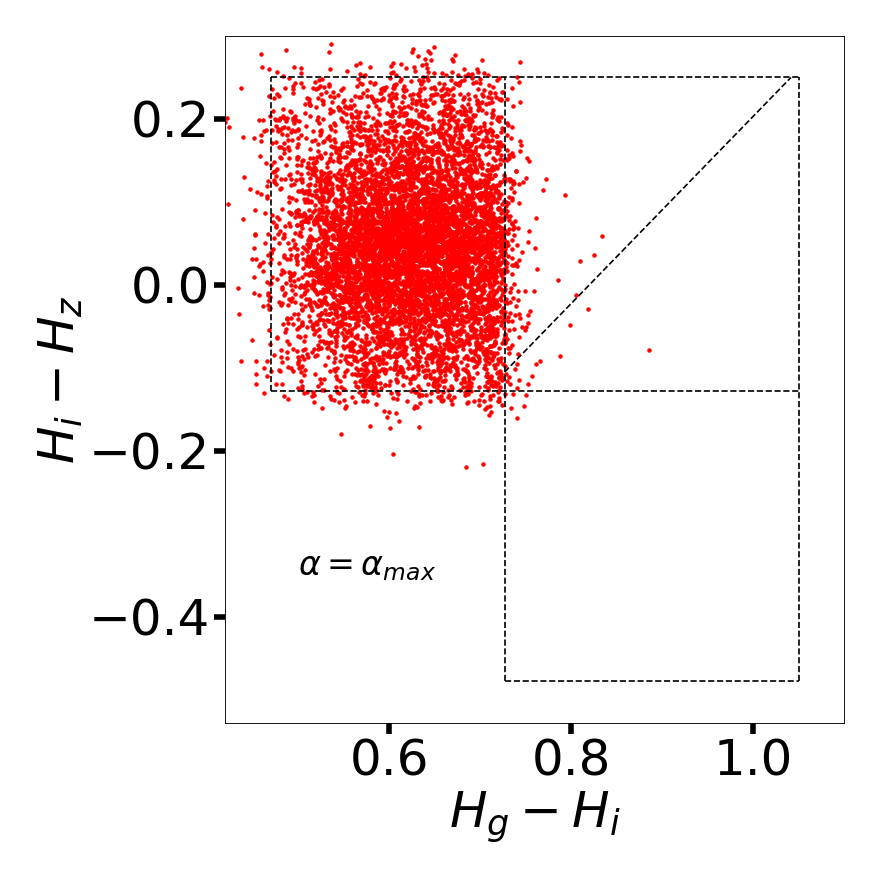}

\caption{Evolution of asteroids within the C-complex at $\alpha=0$ deg with changing phase angle. Left: initial conditions; Middle: $\alpha = \alpha_c$; Right: $\alpha = \alpha_{max}$.}\label{fig:appC}%
\end{figure*}

\begin{figure*}
    
\centering
 \includegraphics[width=4.4cm]{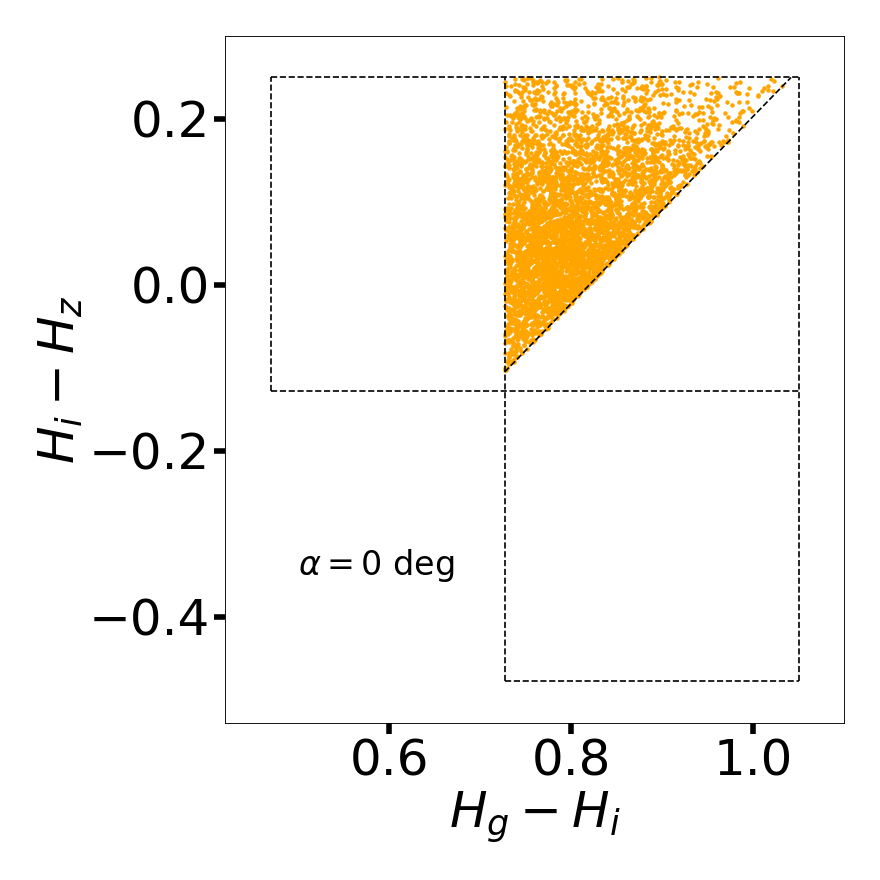}
 \includegraphics[width=4.4cm]{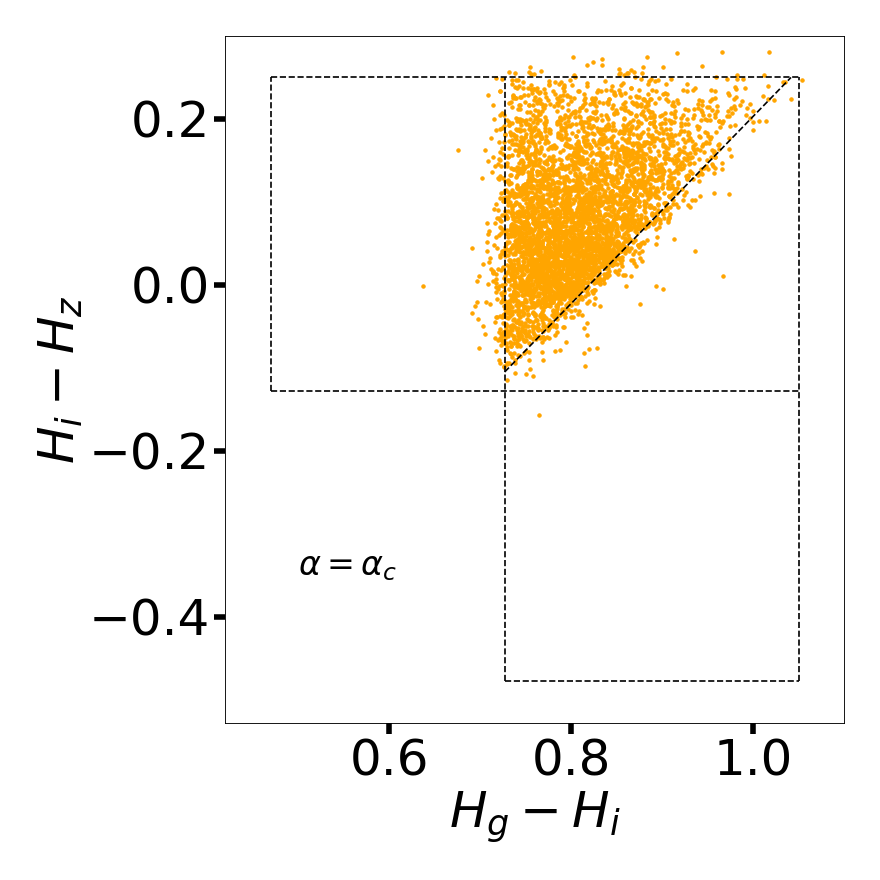}
 \includegraphics[width=4.4cm]{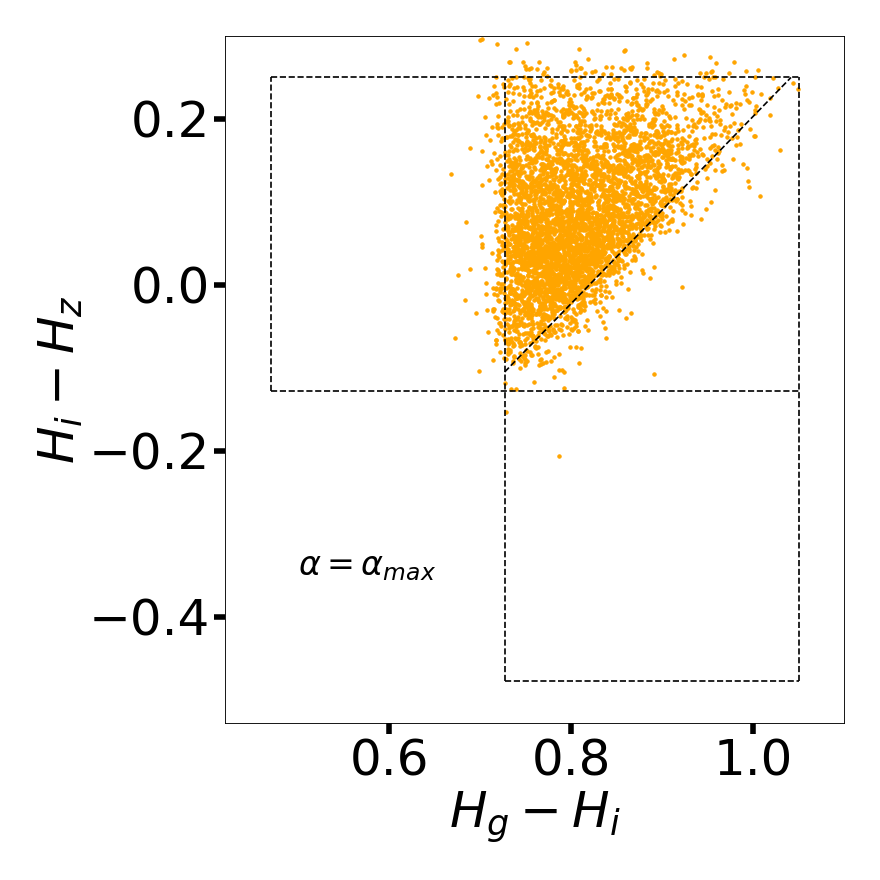}

\caption{Evolution of asteroids within the X-complex at $\alpha=0$ deg with changing phase angle. Left: initial conditions; Middle: $\alpha = \alpha_c$; Right: $\alpha = \alpha_{max}$.}\label{fig:appX}%
\end{figure*}

\begin{figure*}
\centering
 \includegraphics[width=4.4cm]{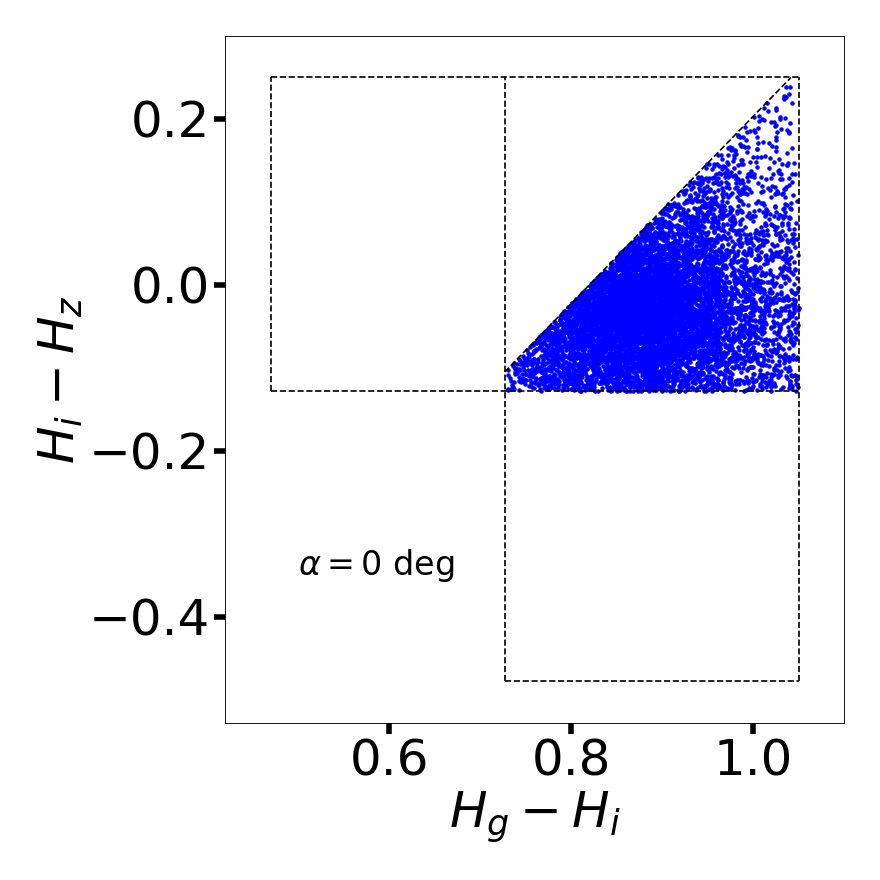}
 \includegraphics[width=4.4cm]{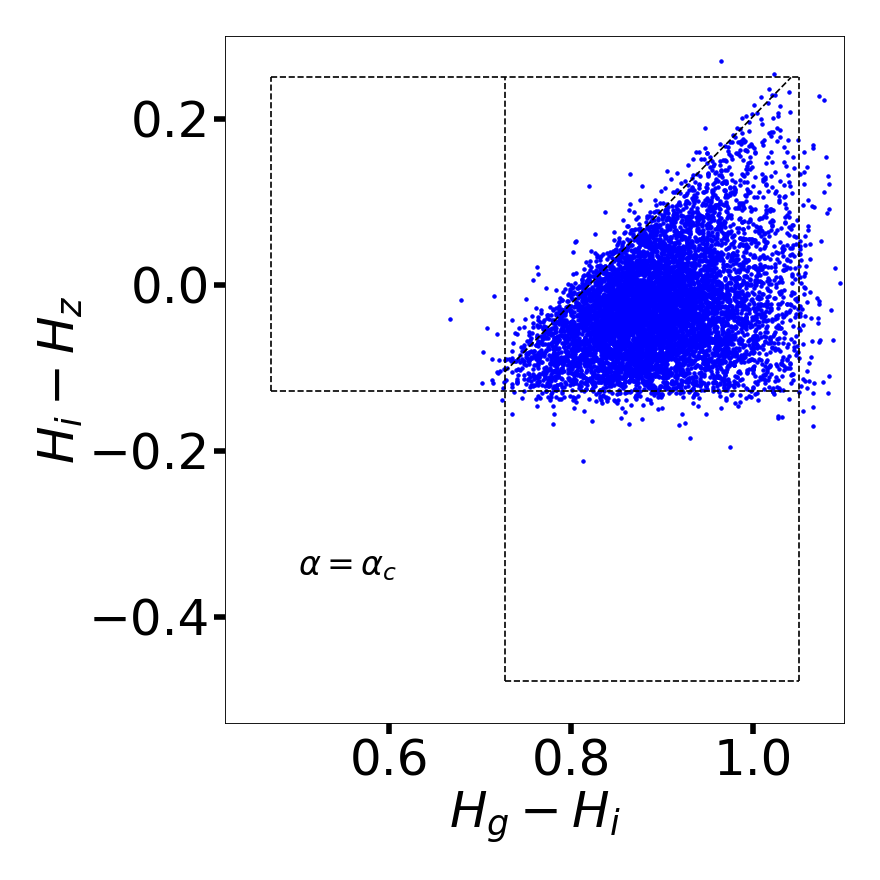}
 \includegraphics[width=4.4cm]{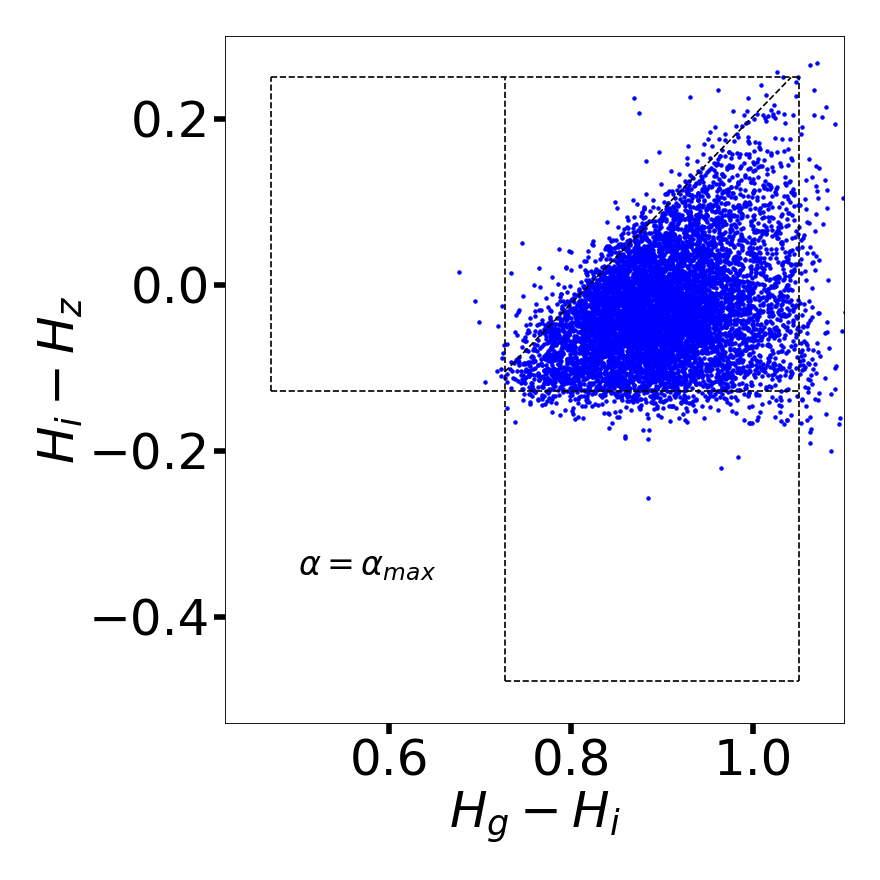}

\caption{Evolution of asteroids within the S-complex at $\alpha=0$ deg with changing phase angle. Left: initial conditions; Middle: $\alpha = \alpha_c$; Right: $\alpha = \alpha_{max}$.}\label{fig:appS}%
\end{figure*}

\begin{figure*}
\centering
 \includegraphics[width=4.4cm]{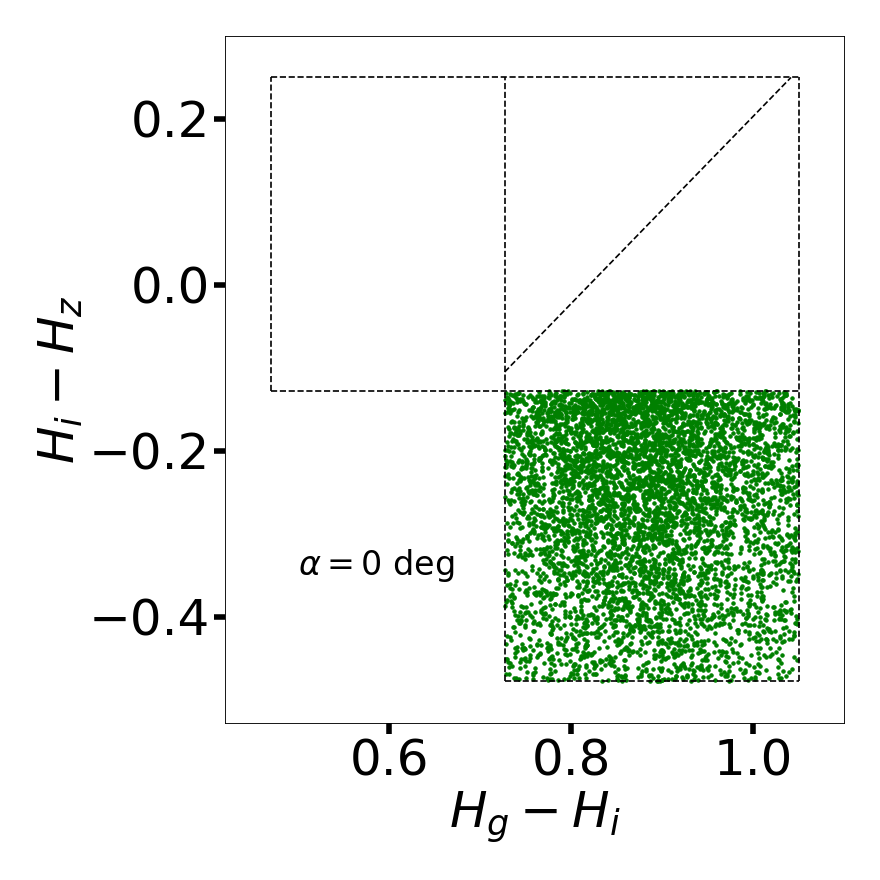}
 \includegraphics[width=4.4cm]{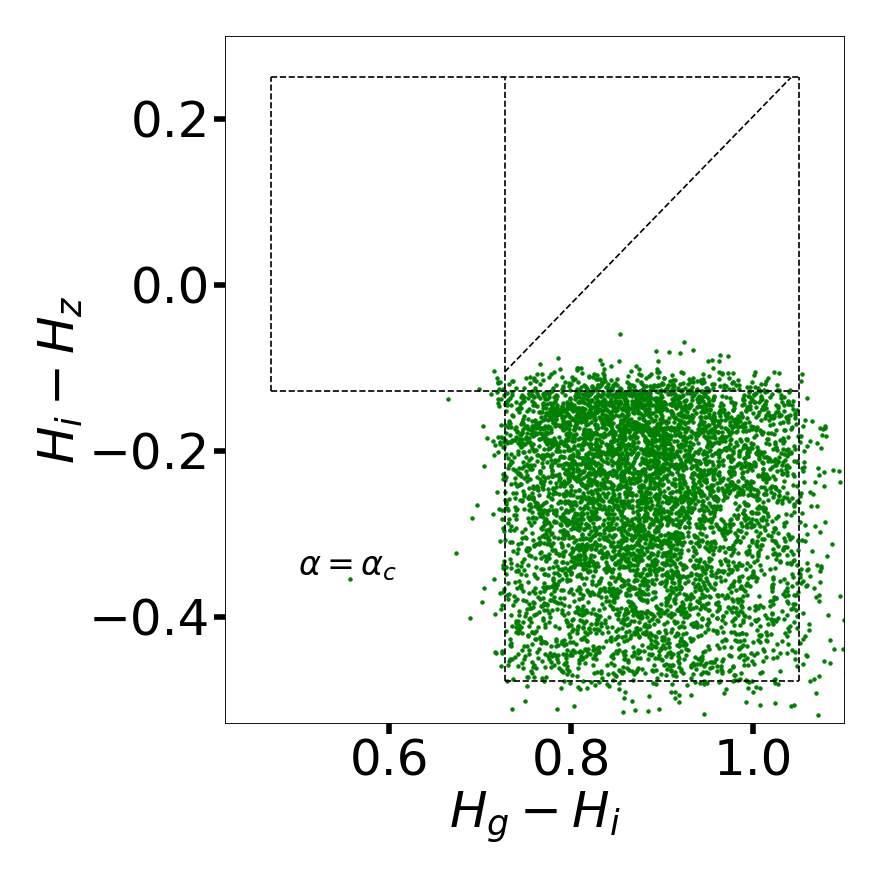}
 \includegraphics[width=4.4cm]{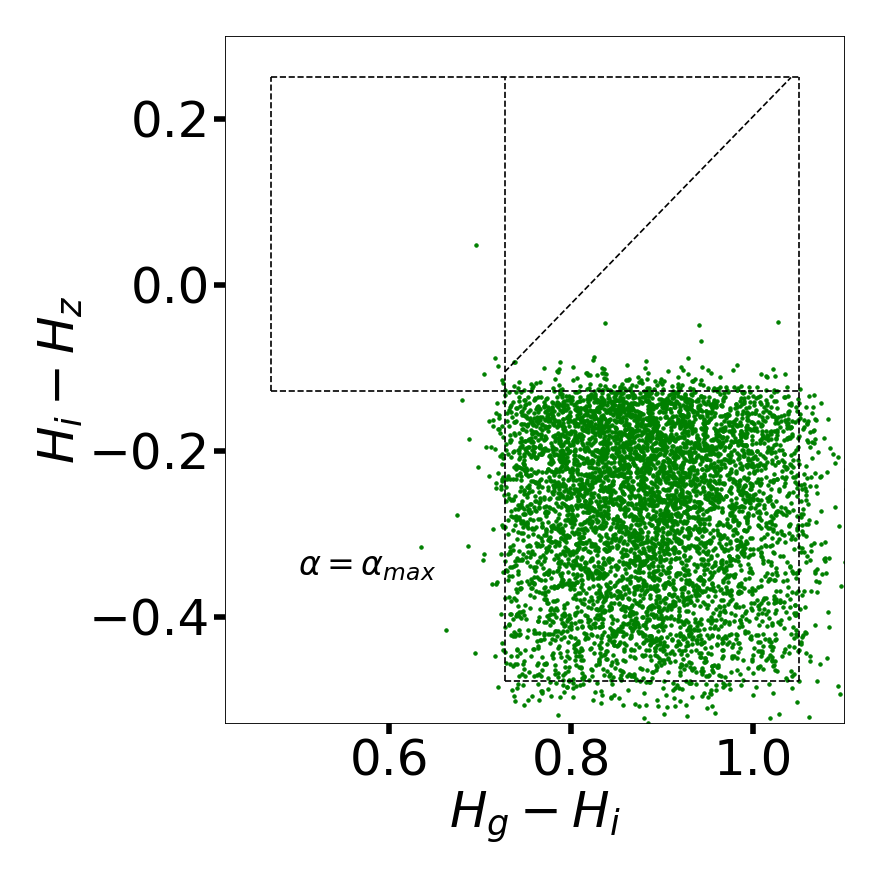}

\caption{Evolution of asteroids within the V-complex at $\alpha=0$ deg with changing phase angle. Left: initial conditions; Middle: $\alpha = \alpha_c$; Right: $\alpha = \alpha_{max}$.}\label{fig:appV}%
\end{figure*}

The change in spectral slope and colors, as seen in Sects. \ref{sec:coloring} and \ref{sec:colors} predict that objects may change their taxon due to changes in the $\alpha$. In Sect. \ref{sec:taxa} we made a crude estimate of the fractional changes in four defined boxes, representing the four major complexes: C, X, S, and V. But, as mentioned, there was no attempt to track the behavior of particular objects. In Figs. \ref{fig:appC} to \ref{fig:appV} we show explicitly how objects initially in one of the boxes evolve with changing $\alpha$: {from opposition to $\alpha_c$, and from $\alpha_c$ to $\alpha_{max}=35$ deg.}

It is clear that most {changes in taxonomic assignment comes} in the limiting regions between boxes and that some objects may suffer significant changes, although the bulge of the population stays within the borders. In the figures, any object leaving the regions limited by the boxes is labeled as {\it unclassified}.

\section{The two behaviors}\label{appB}
{In Sect. \ref{sec:coloring} we detected that the slopes of the asteroids follow two different behaviors (in two different regimes): the BR group first becomes bluer until $\alpha_c$, to start getting redder for larger $\alpha$. In contrast, the opposite holds for the RB group. In this appendix, we explore how both groups behave in the face of the different plots shown in the main text, namely Figs. \ref{fig:szerovss_alpha} to \ref{fig:avgi}.

We analyze first the behavior of the change of the $S^{\prime}_{\alpha}$ with respect to $S0$. In this case, we separated in blue and red dots according to whether their behavior was to go blue or red, according to Fig. \ref{fig:slope0}. Note that in this case, we use, as in Fig. \ref{fig:szerovss_alpha}, the total probability distributions of $H$s and $G$s to understand the impact of the observational results better.
\begin{figure*}
\centering
 \includegraphics[width=4.4cm]{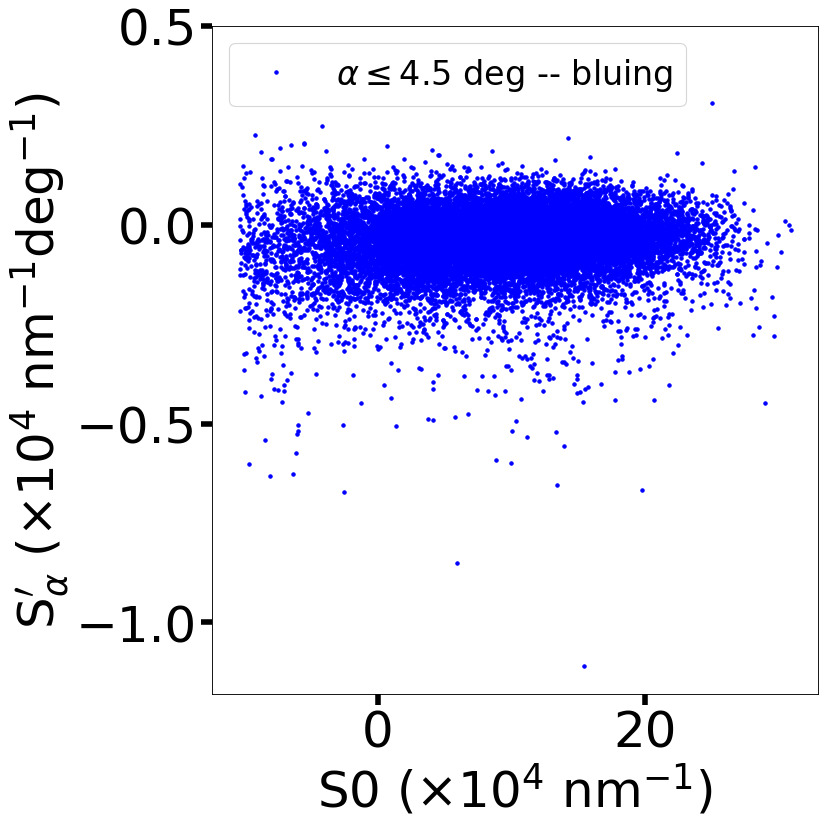}
 \includegraphics[width=4.4cm]{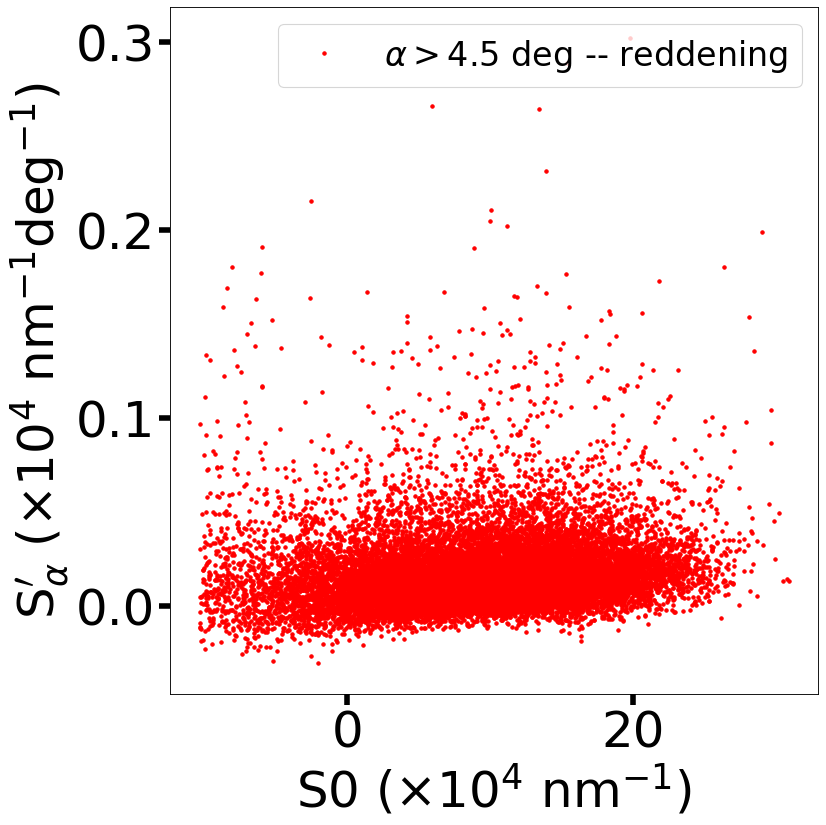}
 \includegraphics[width=4.4cm]{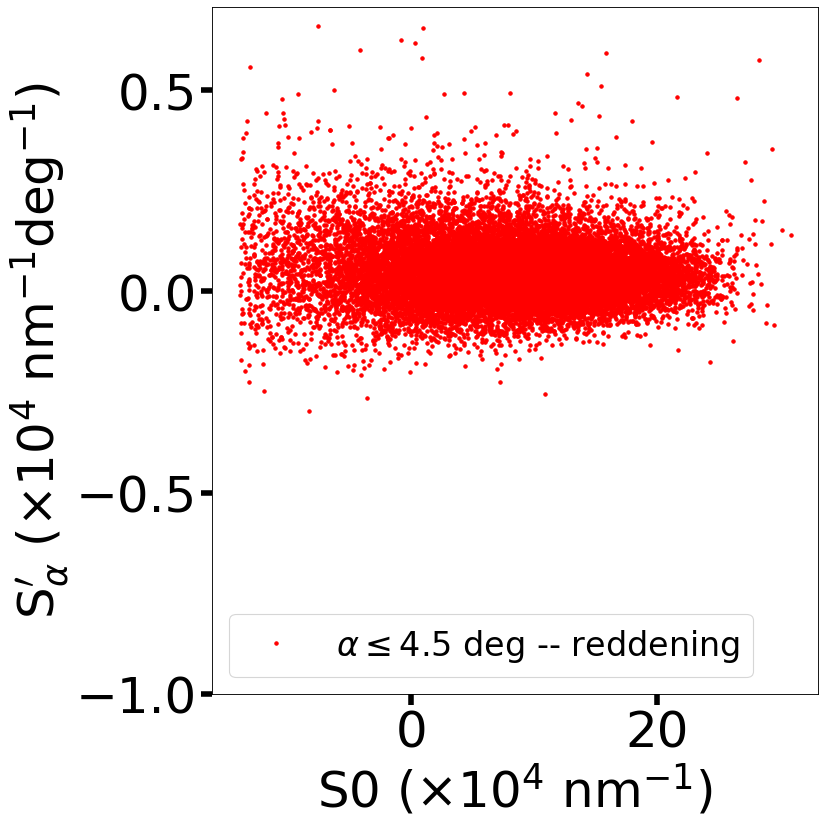}
 \includegraphics[width=4.4cm]{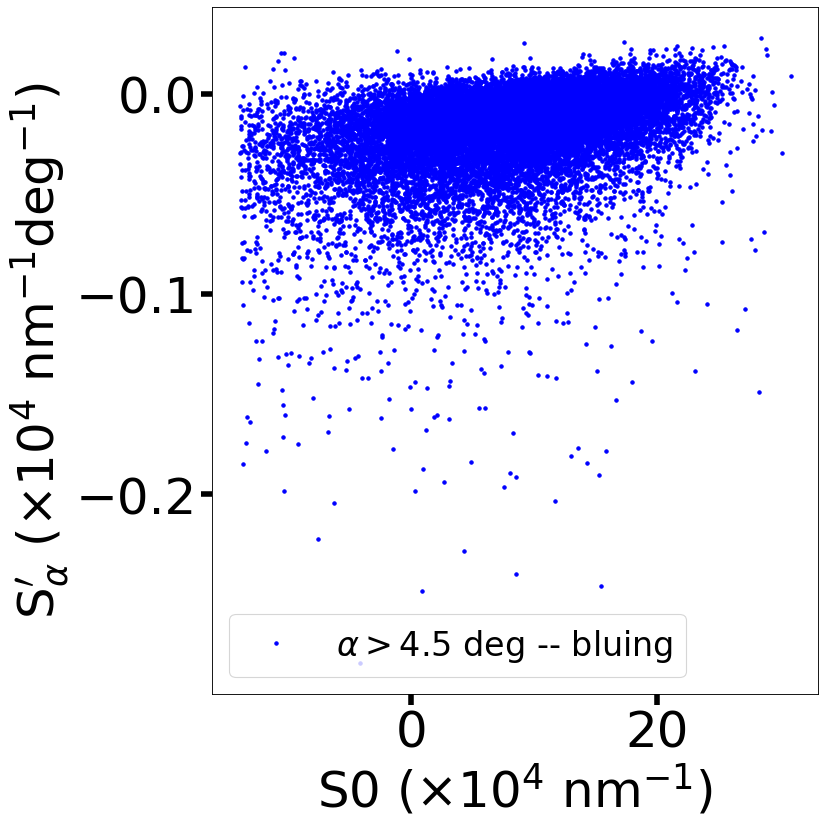}
\caption{$S^{\prime}(\alpha)$ versus $S0$. Left panels: BR group. Right panels: RB group.}\label{fig:appB1}%
\end{figure*}
For both groups, the behavior is the same at $\alpha>\alpha_c$: intrinsically redder objects tend to increase their slope faster than intrinsically bluer objects. In both cases, the Spearman rank order test predicts correlated quantities with p-values virtually equal to zero. On the other hand, the slopes below the critical angle tend to increase with $S0$ for the BR group, while they seem to decrease slightly for the RB group. In both cases, the Spearman index $r_S\leq |0.1|$.

The following figure (Fig. \ref{fig:appB2} is analogous to Fig. \ref{fig:avslopes}) shows the average slope changes in different bins of the semi-major axis (as discussed in the text). In this case, and as expected, we detect both behaviors when separating the BR and RB groups. However, the scales of the average changes are more significant than when considering all objects together. This suggests that the effect tends to cancel out when including all objects, with a slight dominance of the RB group.
\begin{figure*}
\centering
 \includegraphics[width=4.4cm]{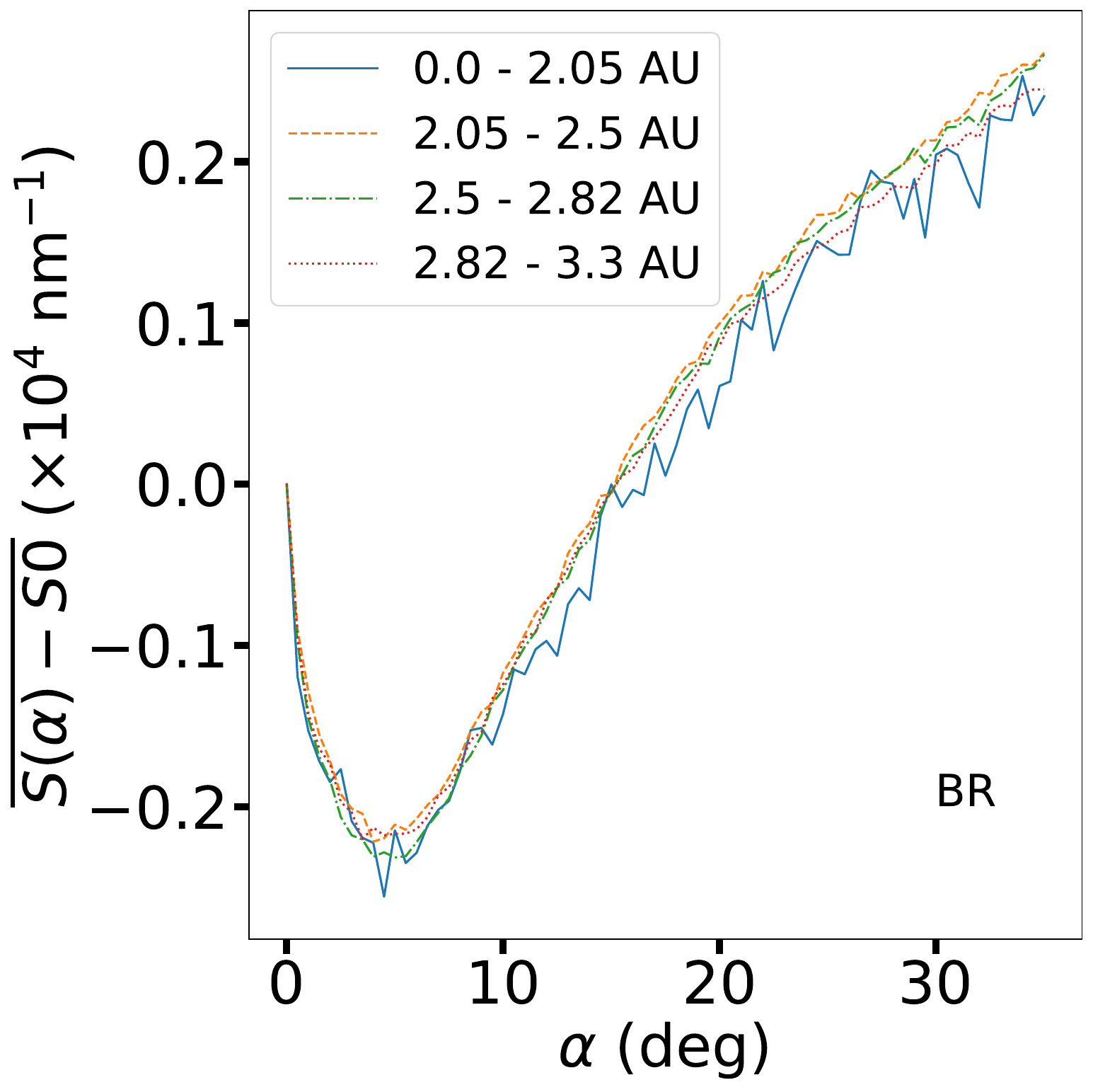}
 \includegraphics[width=4.4cm]{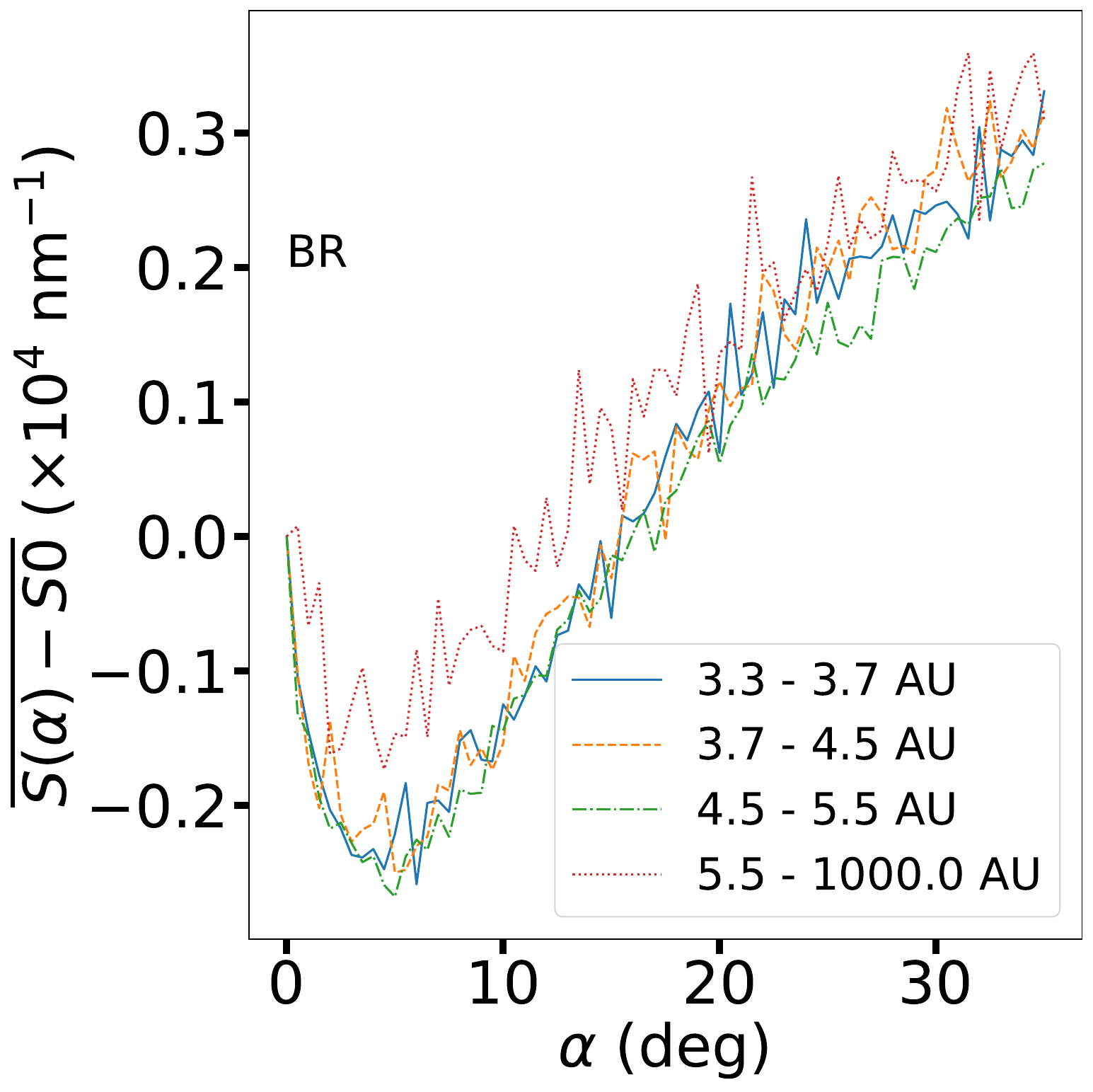}
 \includegraphics[width=4.4cm]{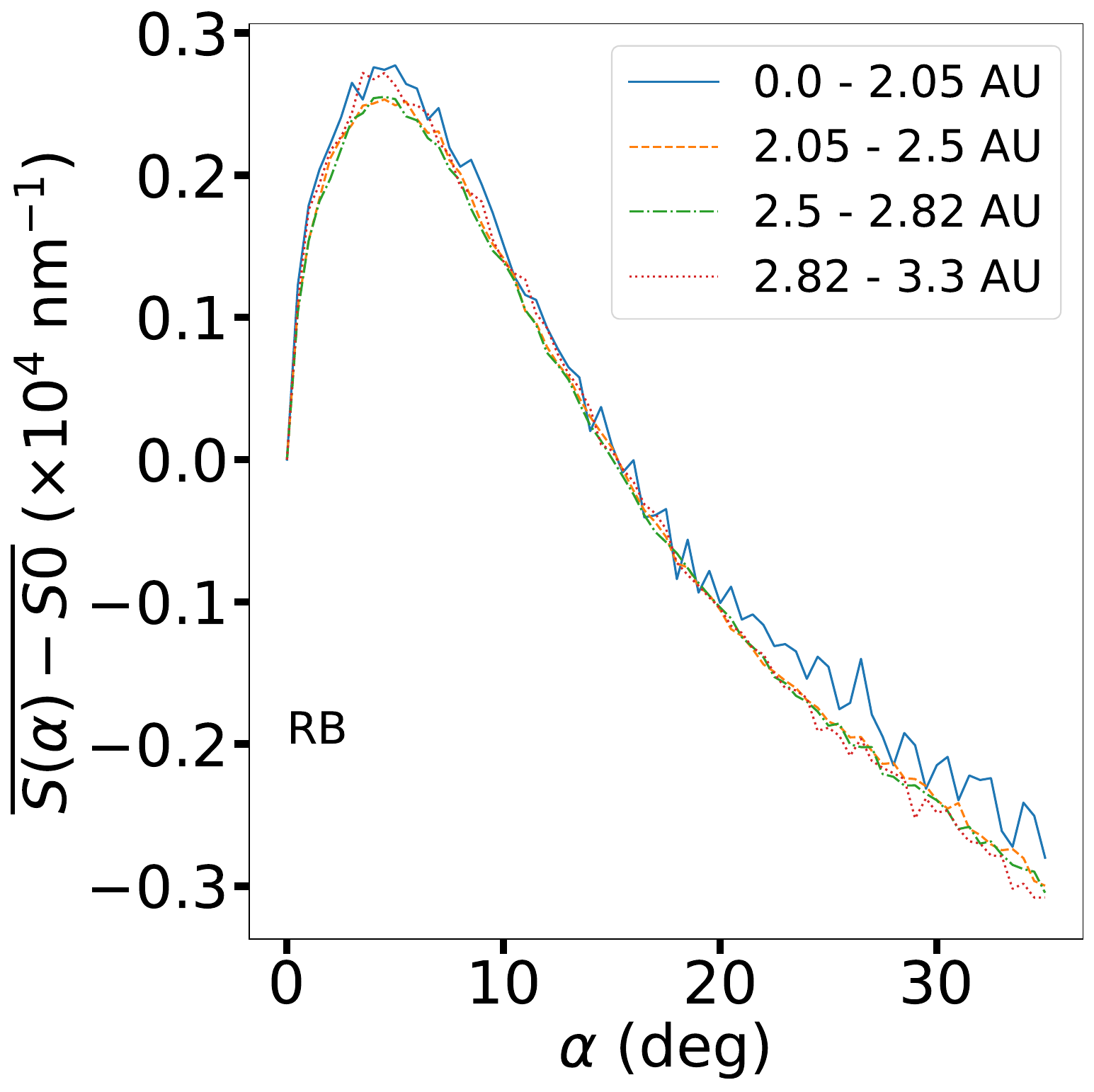}
 \includegraphics[width=4.4cm]{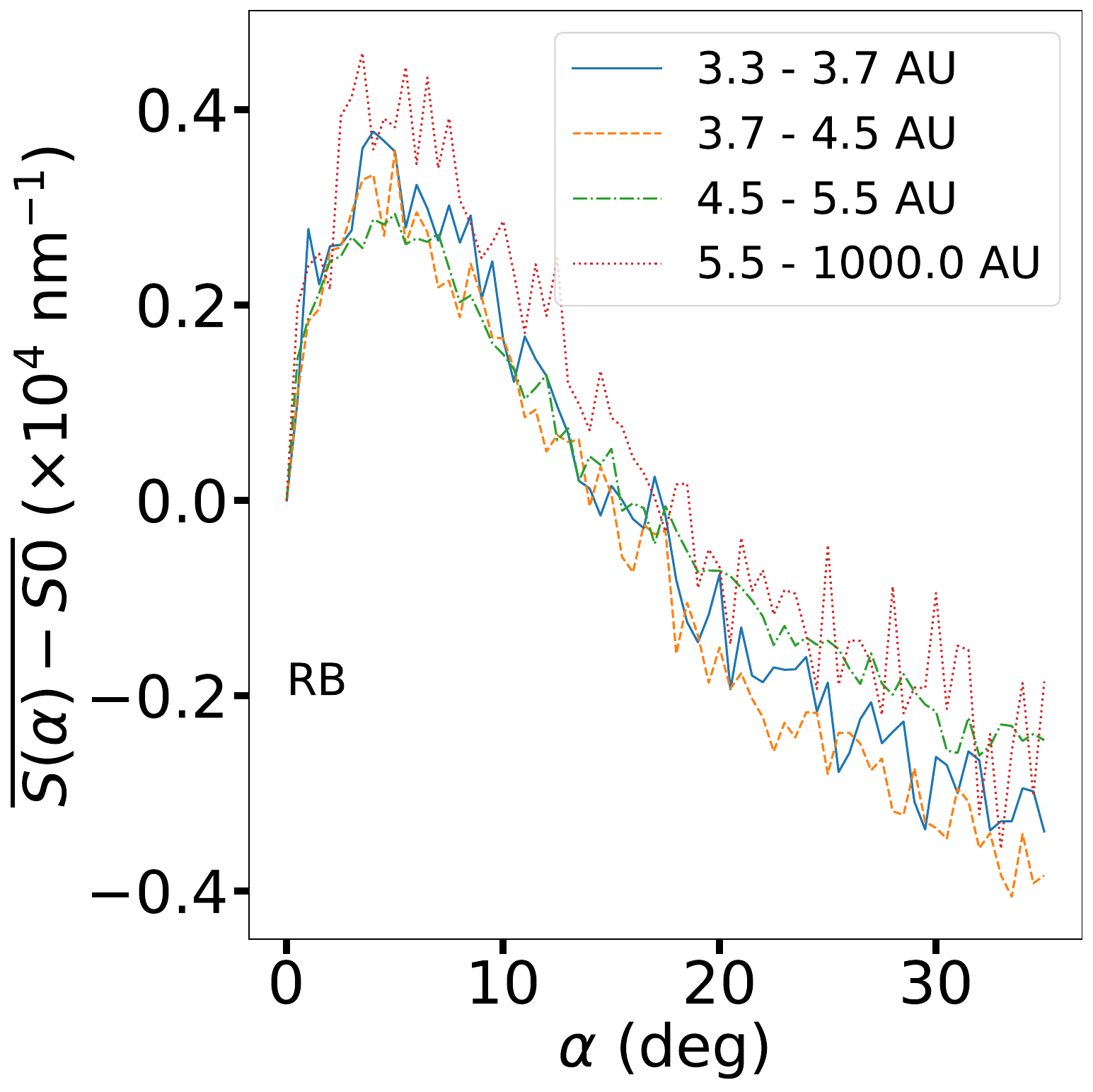}
\caption{Average slope for different semi-major axis bins. Left panels: BR group. Right panels: RB group.}\label{fig:appB2}%
\end{figure*}

Regarding the colors $H_i-H_z$ and $H_g-H_i$, we see that the latter correlates with the spectral slopes, as expected (Fig. \ref{fig:appB3}). Perhaps the most exciting difference comes from the former color, which changes its overall behavior according to whether the objects are from the RB or the BR groups. The BR objects first increase the color and then decrease it, while the opposite is seen for the RB group (Fig. \ref{fig:appB4}); in a way, the color anti-correlates with the change in slope or $H_g-H_i$.
\begin{figure*}
\centering
 \includegraphics[width=4.4cm]{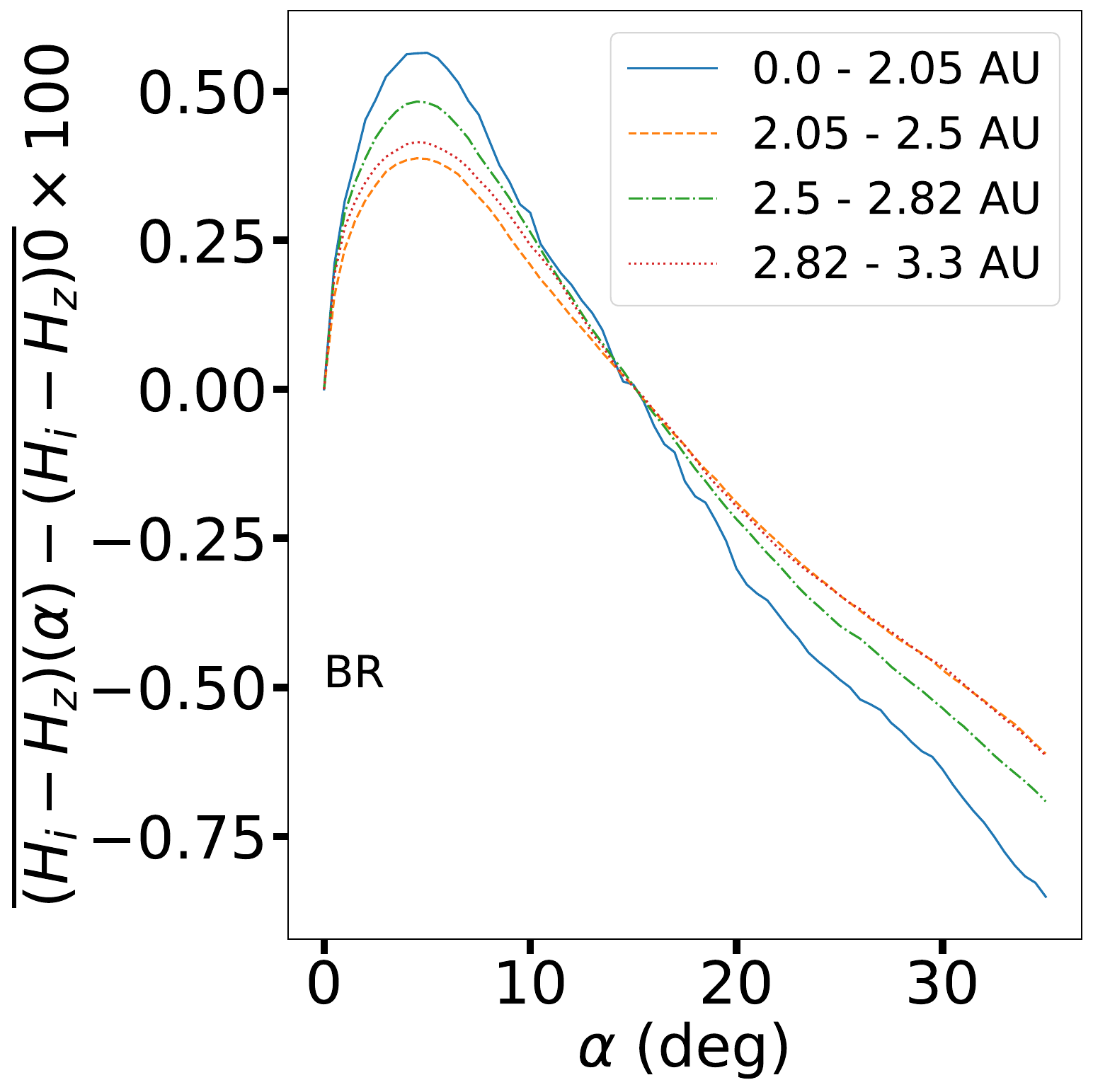}
 \includegraphics[width=4.4cm]{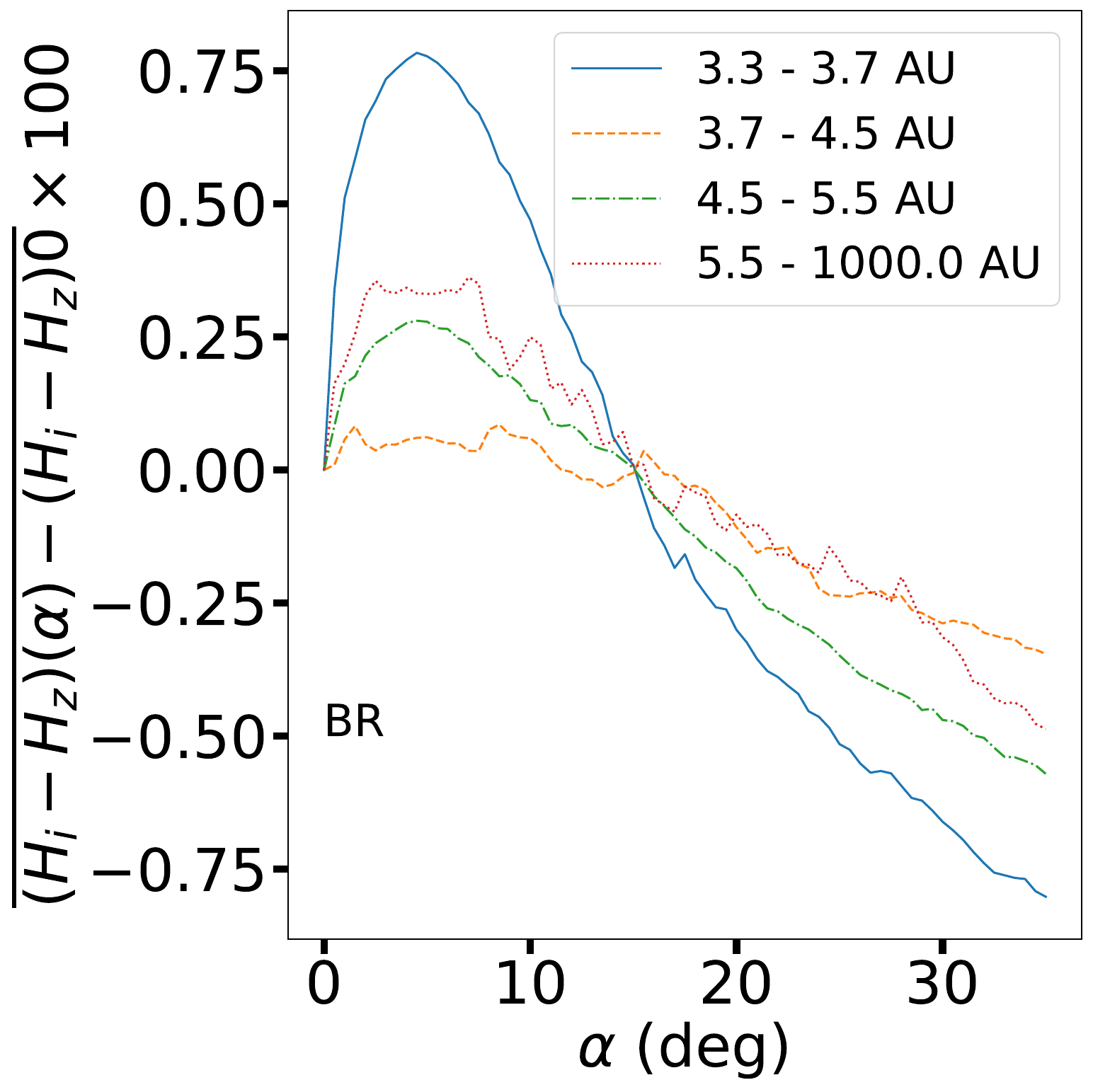}
 \includegraphics[width=4.4cm]{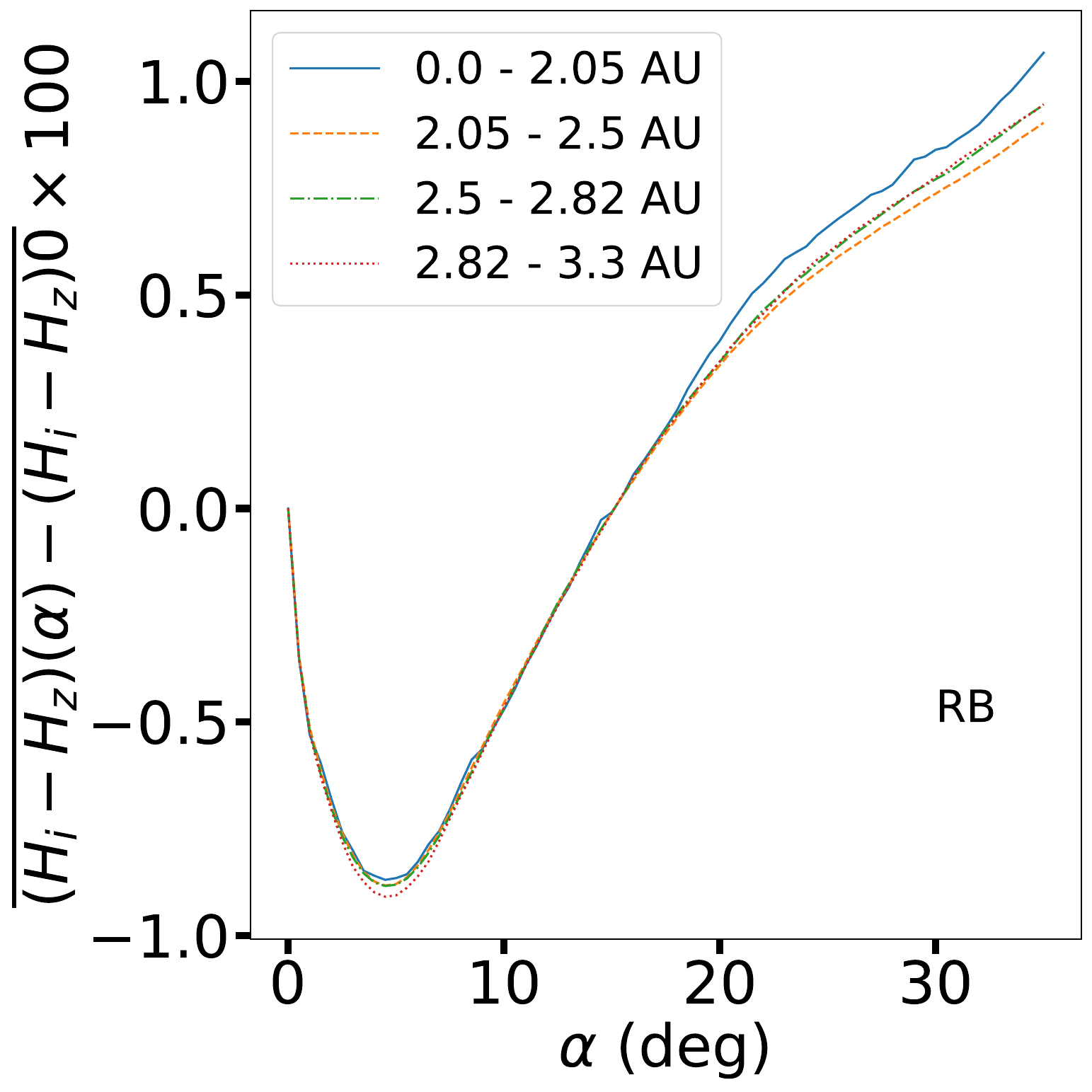}
 \includegraphics[width=4.4cm]{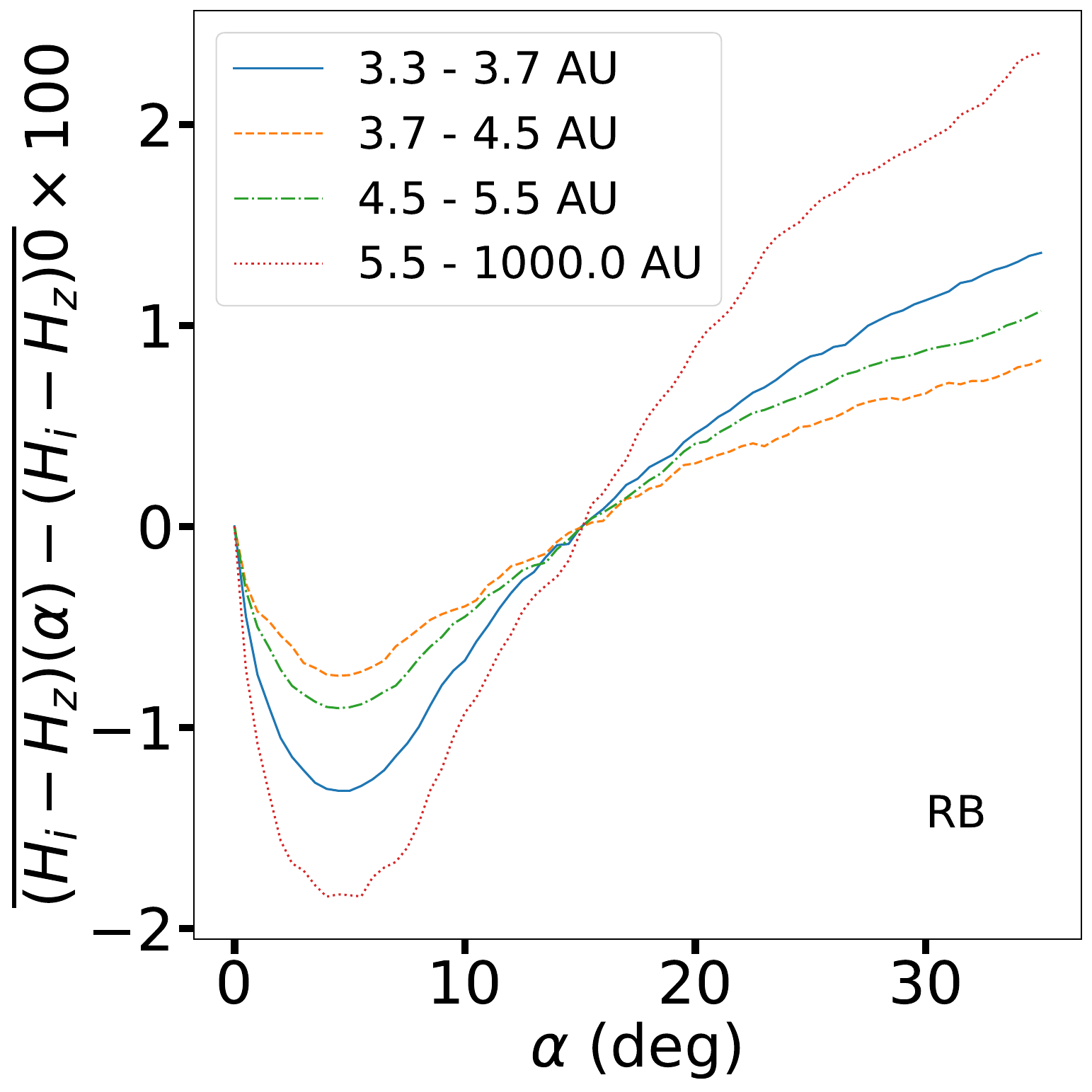}
\caption{Average $H_i-H_z$ versus $\alpha$. Left panels: BR group. Right panels: RB group.}\label{fig:appB3}%
\end{figure*}
\begin{figure*}
\centering
 \includegraphics[width=4.4cm]{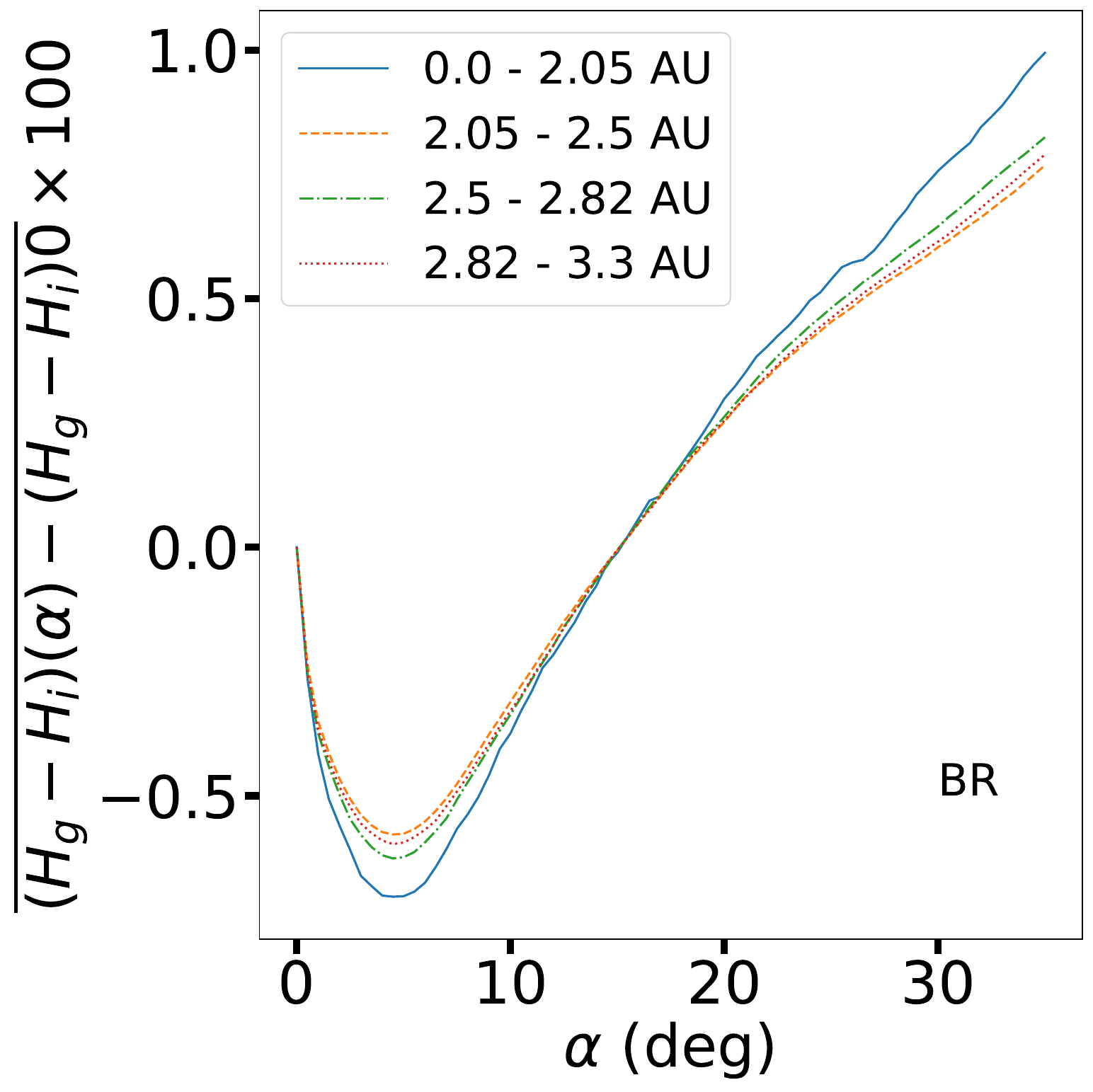}
 \includegraphics[width=4.4cm]{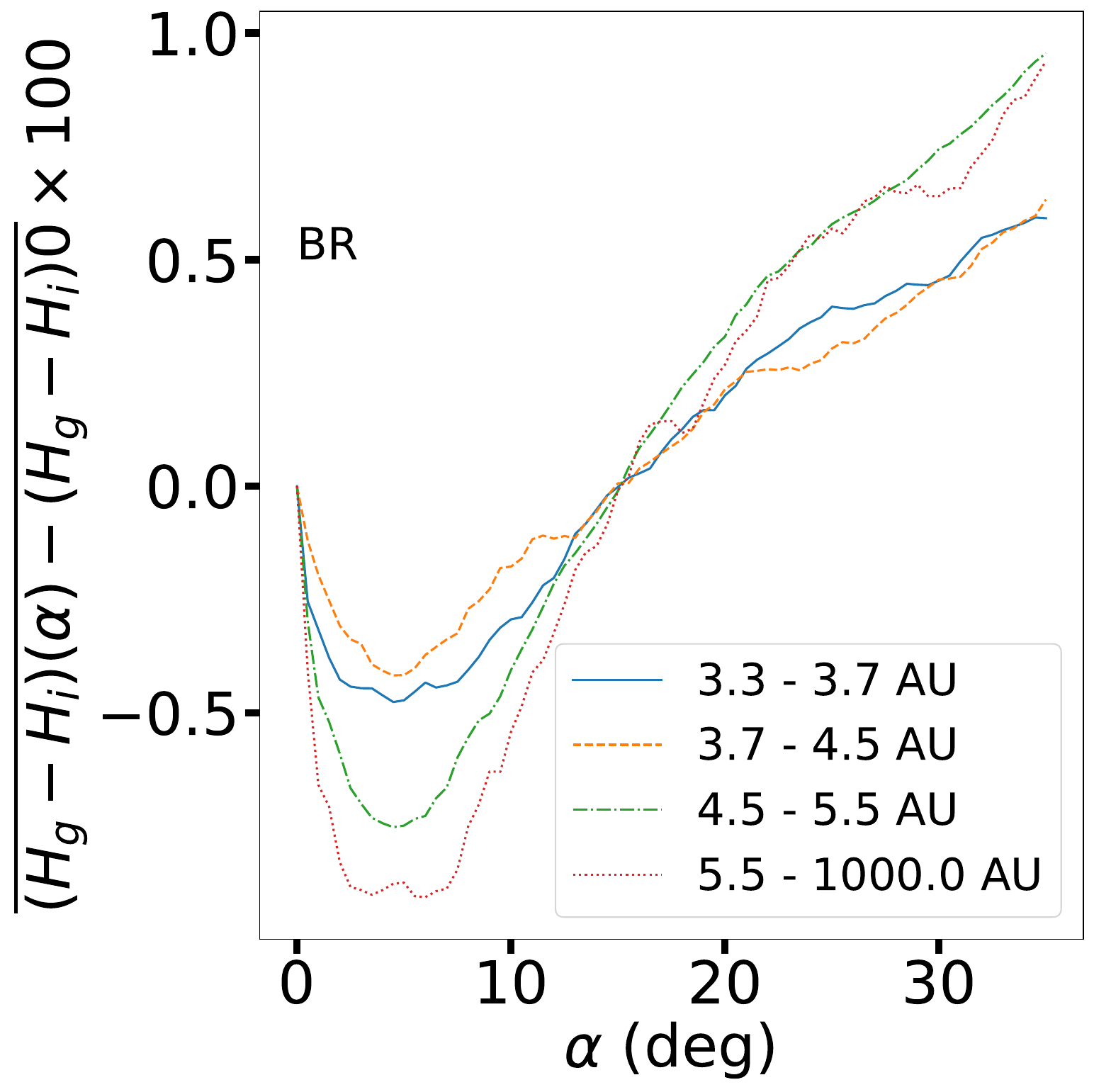}
 \includegraphics[width=4.4cm]{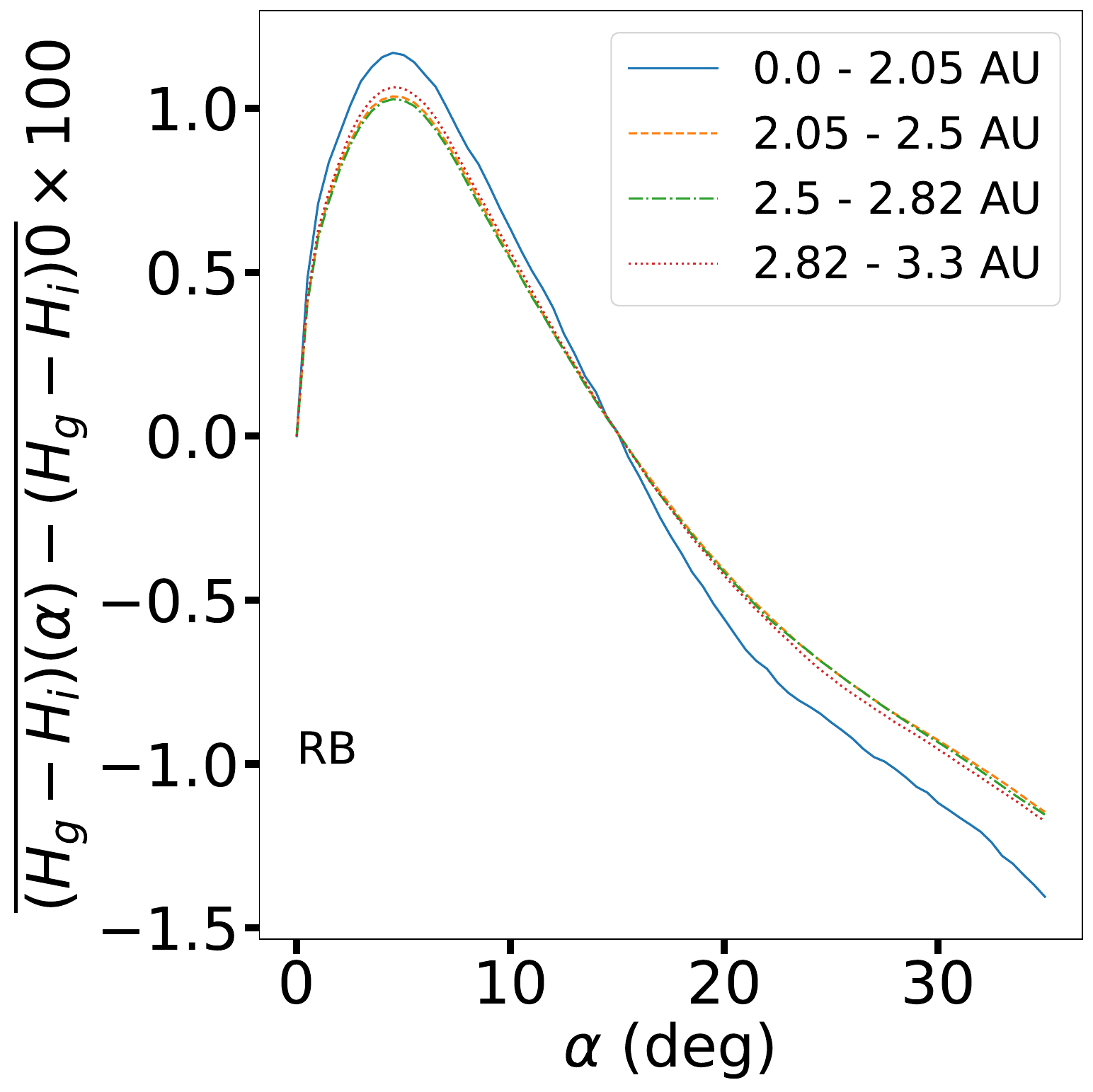}
 \includegraphics[width=4.4cm]{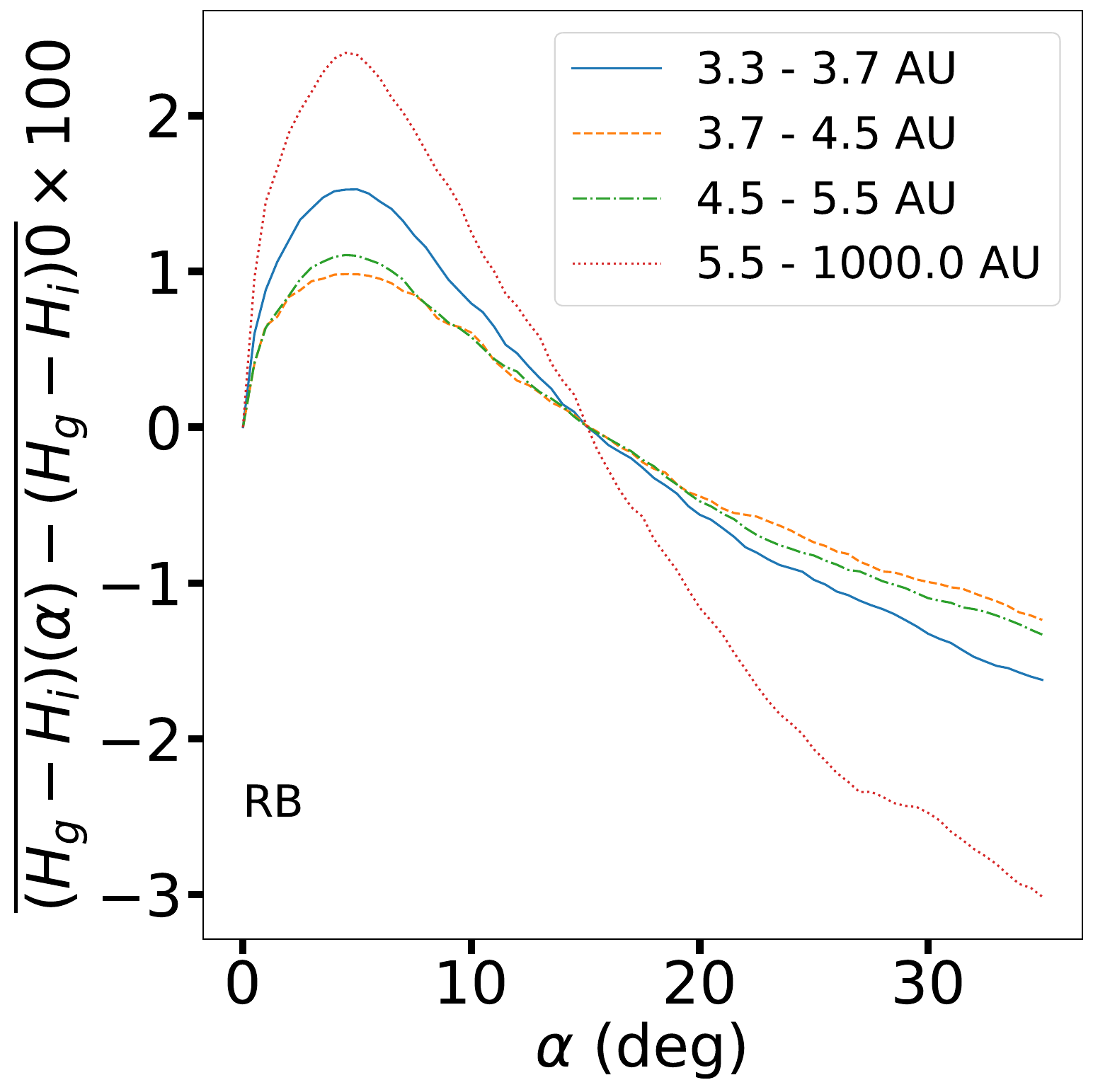}
\caption{Average $H_g-H_i$ versus $\alpha$. Left panels: BR group. Right panels: RB group.}\label{fig:appB4}%
\end{figure*}

The results shown for the colors indicate, on average, that, for the BR objects, the photo-spectra first becomes flatter and perhaps slightly convex, to then go a bit concave. While the RB group goes in the other direction.}

\end{appendix}

%
%
 \bibliographystyle{aa}
 \bibliography{paper}

\end{document}